\documentclass[twocolumn,apj,usenatbib,usegraphicx,useepstopdf]{emulateapj}
\usepackage{color}
\usepackage{comment}
\usepackage{amsmath}
\usepackage{hhline}
\usepackage{CJK}
\usepackage{morefloats}
\usepackage{graphicx}
\usepackage[toc]{appendix}
\usepackage{epstopdf}  
\usepackage{natbib}

\def\mbf#1{\mbox{\boldmath ${#1}$}}
\def\Alfven{Alfv\'{e}n~}
\def\Alfvenic{Alfv\'{e}nic~}

\shorttitle{Outflows by \Alfvenic flux to Fermi Bubbles}
\shortauthors{Suzuki \& Lazarian}

\begin{document}

\title{Galactic Outflows by Alfv\'{e}nic Poynting Flux: Application to Fermi Bubbles}

\author{Takeru K. Suzuki$^{1,2}$ \& Alex Lazarian$^3$}
\email{stakeru@ea.c.u-tokyo.ac.jp}
\altaffiltext{1}{School of Arts \& Sciences, the University of Tokyo,
3-8-1, Komaba, Meguro, Tokyo 153-8902, Japan; }
\altaffiltext{2}{
Department of Physics, Nagoya University, Furo-cho, Chikusa,
  Nagoya, Aichi 464-8602, Japan 
}
\altaffiltext{3}{Department of Astronomy, University of Wisconsin-Madison,
  2535 Sterling Hall, 475 North Charter Street, Madison, WI 53706-1507, USA}



\begin{abstract}
  We investigate roles of magnetic activity in the Galactic bulge region in
  driving large-scale outflows of size $\sim 10$ kpc.
  Magnetic buoyancy and breakups of channel flows formed by magnetorotational
  instability excite Poynting flux by the magnetic tension force.
  A three-dimensional global numerical simulation shows that the average
  luminosity of such \Alfvenic Poynting flux is $10^{40} - 10^{41}$ erg s$^{-1}$.
  We examine the energy and momentum transfer from the Poynting flux to the gas
  by solving time-dependent hydrodynamical simulations with explicitly
  taking into account low-frequency \Alfvenic waves of period of 0.5 Myr
  in a one-dimensional vertical magnetic flux tube. 
  The \Alfvenic waves propagate upward into the Galactic halo, and they are
  damped through the propagation along  meandering magnetic field lines.
  If the turbulence is nearly trans-Alfv\'{e}nic, the wave damping is
  significant, which leads to the formation of an upward propagating shock wave.
  At the shock front, the temperature $\gtrsim 5\times 10^6$ K, the
  density $\approx 6\times 10^{-4}$ cm$^{-3}$, and the outflow velocity
  $\approx 400-500$ km s$^{-1}$ at a height $\approx 10$ kpc, which 
  reasonably explain the basic physical properties of the thermal component
  of the Fermi bubbles.   
\end{abstract}

\section{Introduction}
\label{sec:intro}
Large-scale outflows from the Galactic center (GC hereafter) region are
observed in various wavelengths. Radio continuum observations identified
the North Polar Spur (NPS hereafter), which extends above the Galactic
latitude, $b > 60$ degrees \citep{ber71}.
Although it might be interpreted by local supernova remnants \citep{ber71,ea95},
\citet{sof77,sh84} claimed that this spur is linked to the GC.
Later, large-scale bipolar structures
were detected by X-ray observations \citep{sno97,sof00}, although their
origin was still vague.
\citet{bh03} finally identified bipolar structures and winds directly
originating from the GC at the footpoint regions of these large-scale
structures at mid-infrared wavelengths. 
Based on this observation, they further evaluated the energetics of the
large-scale Galactic outflows from the GC in a quantitative sense,
taking into account angular projection effect precisely \citep[see][for review]{vei05}.

Large-scale ``haze'' with the size of several kpc is also detected toward the GC
in microwaves by WMAP \citep{fin04,df08} and Planck \citep{pla13}. 
Recently, the Fermi-LAT detected large-scale bipolar structures from the GC
covering up to $b \approx \pm 50$ degrees or $\pm 10$ kpc \citep{dob10,su10},
which are called Fermi bubbles.
Emission mechanisms of the non-thermal $\gamma$-rays from the Fermi bubbles are
discussed from hadronic \citep{ca11,fuj13} and leptonic \citep{ms11,che15}
processes. 

The origin of these bipolar structures is still under debate; 
a proposed mechanism is the past activity of the super-massive blackhole
(SMBH hereafter) at the GC \citep{zub11,gm12}, which is inferred from X-ray
observations \citep{koy96}; another possibility is winds driven by multiple
supernovae as a result of bursty star formation \citep{cro12,lac14}. 


Quantitative properties of the Fermi bubbles have been known gradually to date.
If the $\gamma$-rays are from the inverse Compton scattering by nonthermal
electrons, following the leptonic scenario mentioned above, the magnetic
field strength is constrained to be $B=5-20$ $\mu$G from the synchrotron
radiation of the same component of the electrons
\citep[][see also \citealt{car13}]{ack14}. 
\citet{kat13,kat15} derived the temperature of the thermal gas in the bubbles
is $T\approx 3.5\times 10^6$ K from spectral fitting to X-ray observations by
{\it Suzaku}. From X-ray emission strengths of ionized oxygens, \citet{mb16}
estimated $T\approx (4-5) \times 10^6$ K and the density
$n \approx 1\times 10^{-3}$cm$^{-3}$ of the thermal component in the bubbles
and the shells covering the bubbles.
From these values, we can estimate the sound speed,
\begin{equation}
  c_{\rm s} =\sqrt{\frac{k_{\rm B}T}{\mu m_{\rm p}}}
  = 230{\rm km\;s^{-1}}\left(\frac{T}{4\times 10^6{\rm K}}\right)^{1/2}
  \left(\frac{\mu}{0.6}\right)^{-1}, 
  \label{eq:cs}
\end{equation}
and the \Alfven speed, 
\begin{eqnarray}
  v_{\rm A} &=& \frac{B}{\sqrt{4\pi \mu m_{\rm p}n}} \nonumber \\
  &=& 890{\rm km\;s^{-1}}\left(\frac{B}{10\mu{\rm G}}\right) \nonumber \\
   & & \hspace{1cm} \left(\frac{n}{10^{-3}{\rm cm}^{-3}}\right)^{-1/2}
   \left(\frac{\mu}{0.6}\right)^{-1/2}, 
  \label{eq:va}
\end{eqnarray}
where $k_{\rm B}$ is the Boltzmann constant and $m_{\rm p}$ is the proton mass;
we here assume the mean molecular weight $\mu=0.6$ as a standard value.
From the comparison between Equations (\ref{eq:cs}) and (\ref{eq:va}), the
magnetic field is expected to affect the dynamics of the bipolar bubbles.
Magnetohydrodynamical (MHD) simulations have been performed in order to
examine roles of magnetic fields in driving bubbles/outflows
\citep{bb14,mou15}.

One of the authors of the present paper also worked on a Global MHD simulation
in the Galactic bulge region \citep{suz15} and found that outflows are driven
by the gradient of magneto-turbulent pressure and magnetic buoyancy
(Parker instability; \citealt{pak66}) in a stochastic manner
\citep[see also][]{mac09,mac13}.
Because the main focus of \citet{suz15} was the generation of random velocities
in the bulge, the simulation box covers up to the elevation angle of $\pm 60$
degrees from the Galactic plane and cannot treat large-scale outflows.
In the present paper, we carry our a pilot study how the Poynting
flux generated from the Galactic Bulge heats up the Galactic halo and drives
outflows, following the energy transfer in a one-dimensional (1D) flux tube
by {\it time-dependent} simulations. 

\citet{bre91} introduced a theoretical framework for the galactic winds driven
by the pressure of thermal gas, cosmic rays, and \Alfven waves under the
{\it time-steady} approximation. This is further extended by explicitly taking into
account specific interaction processes between these three components by
\citet{zir96} and \citet{ptu97}; in particular, they considered
the excitation of \Alfven waves from cosmic rays via streaming instability
\citep{wen68}, and the
dissipation of the excited \Alfven waves by nonlinear Landau damping
\citep{kp69,bl08}, which finally heats up the gas.
The typical wavelength of the \Alfven waves is the gyroradius of relativistic
ions, $r_{\rm g}\approx 3\times 10^{11}$cm$\left(\frac{B}{10\mu{\rm G}}
\right)^{-1}{\gamma}$, where $\gamma$ is the Lorentz factor \citep{kp69,eve08}.
Namely the wavelength is shorter than the pc scale in general.

In the present paper, while we also consider \Alfvenic Poynting flux,
we present a model that is different from the previous works both in terms
of the source of the \Alfvenic waves and the damping mechanism. In particular,
we consider the Poynting flux that is naturally produced by large-scale
magnetic activity driven for example by magnetic loops and use the numerical
simulations in \citet{suz15} for our quantitative studies.
The typical wavelength of these \Alfvenic waves is $\sim 0.1-1$ kpc.
Such long-wavelength \Alfvenic waves hardly suffer nonlinear Landau damping.
Therefore, we cannot assume instantaneous dissipation of the \Alfven waves, which
is often adopted \citep[e.g., ][]{eve08}. Instead, we explicitly treat the
transfer of the \Alfvenic Poynting flux.
For the damping of \Alfvenic waves
we use models of turbulent damping of \Alfvenic waves \citep{yl02,fg04,laz16}.
In these models MHD turbulence efficiently cascades \Alfven waves in the MHD regime, independently of the plasma parameters of the media, but
depending on the level of background turbulence.

It is a key to properly model the damping rate of the \Alfvenic waves and
to calculate both heating and wind launching by the wave damping in a
self-consistent manner. 
If \Alfvenic waves are damped near the Galactic bulge where the density is
high, the final wind velocity is considered to be not fast enough to escape
from the Galaxy. This is first because the momentum injection near the wind base
does not launch an outflow, but lifts up the gas by contributing to the
pressure balance, and second because the energy injection in the high-density region is mainly lost by radiation cooling and does not lead to a substantial
increase of the temperature. On the other hand, wave dissipation
at high latitudes is expected to drive the wind and the velocity of the wind
depends on the density where the wave dissipation mainly happens.
In addition, heating in rarefied gas can significantly increase the
temperature due both to lower mass to be heated and lower cooling efficiency.

Our focus is on the thermal component of the bipolar bubbles and outflows.
Here we briefly describe the global energetics, following \citet{mb16},
in which each bubble is modeled by a simple prolate ellipsoid
with semi-major axis $=$ 5 kpc and semi-minor axis $=$ 3 kpc. The total energy
in one bubble is
\begin{eqnarray}
  E_{\rm th} &=& \frac{k_{\rm B}T}{\mu m_{\rm p}} M_{\rm th} \nonumber \\
  &=& 5.9\times 10^{54}{\rm erg}
  \left(\frac{T}{4\times 10^6{\rm K}}\right)\left(\frac{\mu}{0.6}\right)^{-1}
  \nonumber \\
  & & \hspace{2cm}\left(\frac{M_{\rm th}}{5.4\times 10^6M_{\odot}}\right), 
\end{eqnarray}
where $M_{\rm th}$ is the mass of the thermal component in one bubble. 
If we know the duration, $\tau_{\rm inj}$, of the energy injection from
the bulge, we can estimate the minimum required energy injection rate to keep
$E_{\rm th}$:
\begin{equation}
  L_{\rm inj} = \frac{E_{\rm th}}{\tau_{\rm inj}} = 1.9\times 10^{40}{\rm erg\;
    s^{-1}}\left(\frac{E_{\rm th}}{5.9\times 10^{54}{\rm erg}}\right)
  \left(\frac{\tau_{\rm inj}}{10{\rm Myr}}\right)^{-1}. 
  \label{eq:lumA}
\end{equation}
We should note that $\tau_{\rm inj}$ is very uncertain and the adopted value
in previous works varies from a few $10^5$ Myr \citep[e.g.][]{gm12} to $\gtrsim$ a few $10^8$ Myr
\citep[e.g.][]{cro15} in different theoretical models, whereas too long
$\tau_{\rm inj}\gg$ a few Myr may be unrealistic \citep{bla13}.
Therefore, we leave $\tau_{\rm inj}$ as a free parameter in our model
calculations (Section \ref{sec:model}). 

In Section \ref{sec:3DMHD}, we briefly summarize the global 3D MHD simulation
by \citet{suz15} and examine the Poynting flux ejected from the bulge region
in the numerical data. In Section \ref{sec:model}, we introduce our model to
cover a large vertical region. We present main results
in Section \ref{sec:res} and discuss related topics in Section \ref{sec:dis}

\section{Injection of Alfv\'{e}nic flux from the Galactic Bulge}
\label{sec:3DMHD}
\begin{figure}
  \begin{center}
    \includegraphics[width=0.46\textwidth]{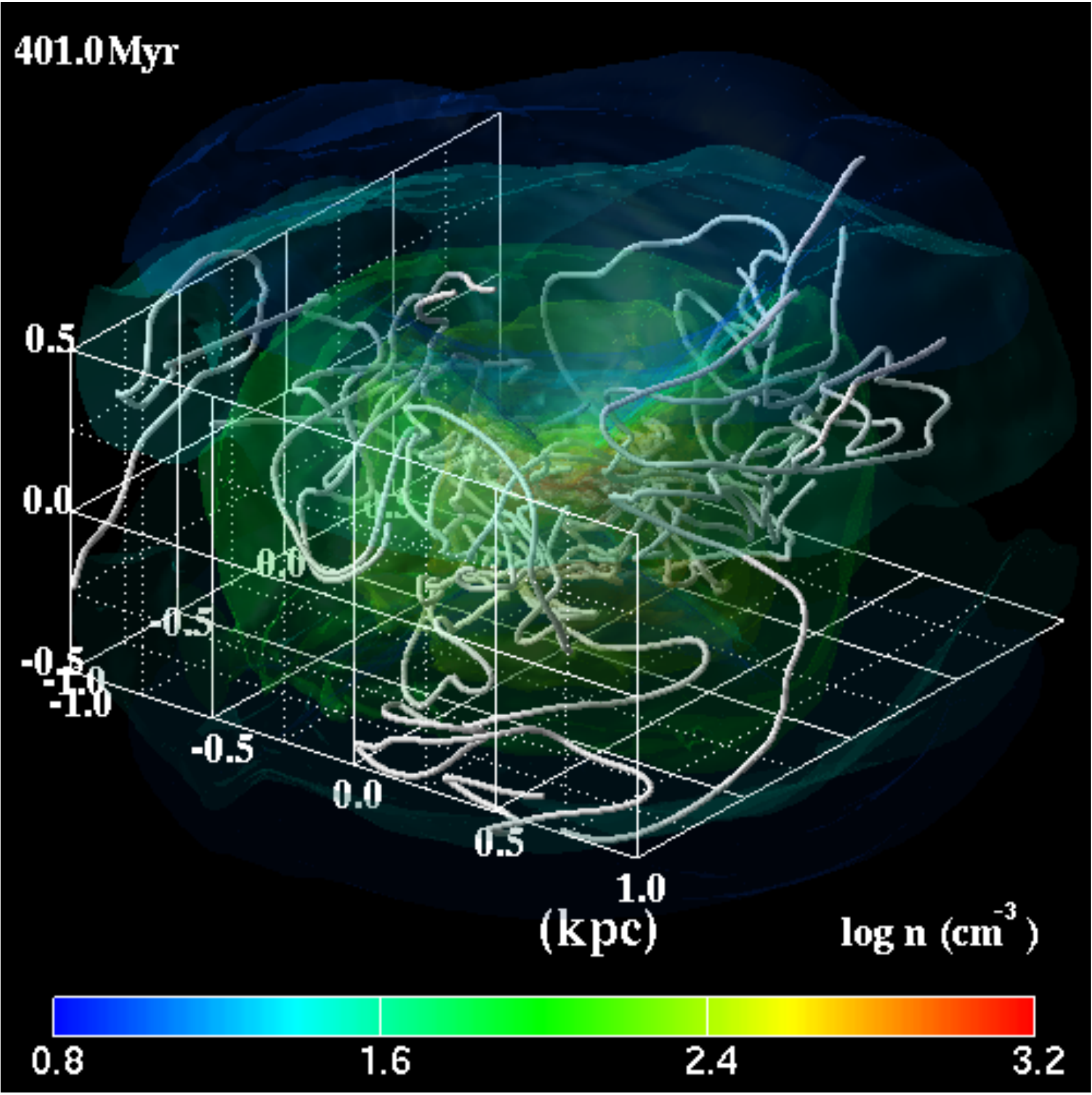}
  \end{center}
  \caption{Snapshot view of magnetic field lines (white lines) and density
    contour (transparent colors) at $t=401.0$ Myr after the simulation 
    \citep{suz15} sets in. 
  \label{fig:snp3D}}
\end{figure}

In this section we briefly describe the global MHD simulation in \citet{suz15}
and quantify the upgoing Poynting flux derived from the simulation, which will
contribute to the heating and driving outflows in the Galactic halo region. 
The simulation was carried out in spherical coordinates, $(r,\theta,\phi)$,
and covers a wide radial region from the inner radius at 0.01 kpc to the
outer boundary at 60 kpc. On the other hand, it covers 
$30^{\circ}\le \theta\le 150^{\circ}$ ($\pm 60^{\circ}$ degrees from the Galactic
plane) and does not treat bipolar outflows ejected from the bulge region.

The simulation started from a very weak vertical magnetic field with $0.7-20$
$\mu$G in the bulge region.
Magnetorotational instability \citep[][MRI hereafter]{vel59,cha61,bh91} triggers
the amplification of the magnetic field by creating the radial component in
high-$\beta$ regions where the gas pressure dominates the magnetic pressure.
At higher altitudes in which the magnetic pressure is comparable to or exceeds
the gas pressure, Parker instability \citep{pak66} also amplifies horizontal
magnetic field to further create the vertical component and excites vertical
flows. The radial differential rotation and vertical shear also amplify
the magnetic field efficiently by field-line stretching \citep[see also][for
Galactic dynamo]{vc01,vs14}.

Figure \ref{fig:snp3D} presents a snapshot of magnetic field lines and density
structure taken from \citet{suz15}\footnote{\citet{suz15} made a mistake when
  converting the nondimensional units used for the simulation to the physical
  units, and the time unit must be smaller by $\approx 10\%$.
  \citet{suz15} presents a snapshot at $t=439.02$ Myr, but this time must be
  corrected to be $t=401.0$ Myr as shown in Figure \ref{fig:snp3D}.}, 
which shows entangled turbulent field lines. 
When the quasi-saturated state is achieved after $t\gtrsim 300$ Myr, the
average field strength in the bulge region is $0.1-1$ mG and it exceeds mG
in local regions with field concentration, which are consistent with the lower
limit $>50$ $\mu$G from $\gamma$-ray observations \citep{cro10} and
an estimated value, $\approx$ a few mG, in a dense cloud \citep{pil15} based
on the \citet{cf53} method. 


\begin{figure}
  \begin{center}
    \includegraphics[width=0.46\textwidth]{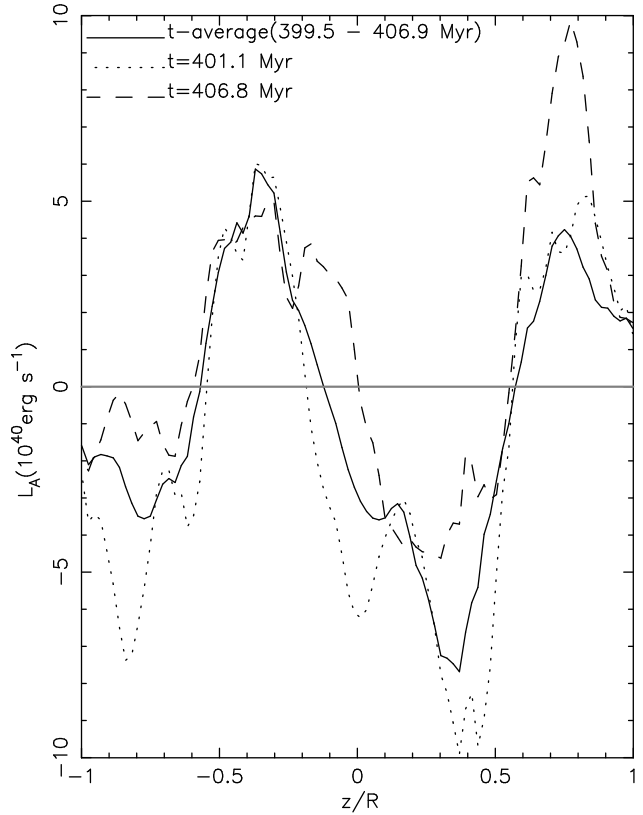}
  \end{center}
  \caption{\Alfvenic luminosity, $L_{\rm A}$, integrated from $R_{\rm in}$($=0.23$ kpc) to $R_{\rm out}$($=1.1$ kpc) as a function of $z/R$. $L_{\rm A}$
    averaged during time $=399.5-406.9$ Myr (solid line) is compared to
    snapshots at $401.1$ Myr (dotted line) and  $406.8$ Myr (dashed line). 
  \label{fig:LA}}
\end{figure}



\begin{figure}
  \begin{center}
    \includegraphics[width=0.46\textwidth]{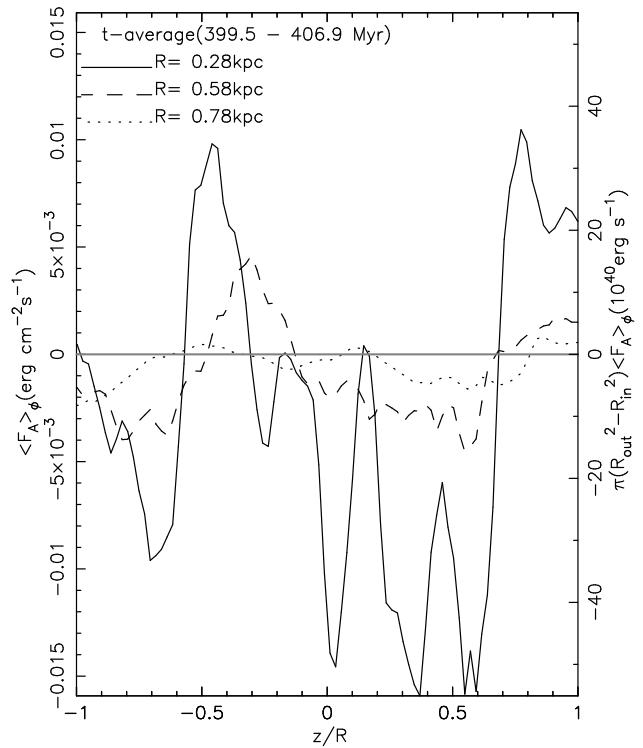}
  \end{center}
  \caption{\Alfvenic energy flux averaged over time and azimuthal angle at
    different $R=0.28$ kpc (solid), 0.58 kpc (dashed), and 0.78 kpc (dotted). 
  \label{fig:FA}} 
\end{figure}

Although \citet{suz15} discussed outflows from the bulge region, they did not
examine the outgoing Poynting flux.  
In order to connect to the 1D vertical flux tube we model in Section
\ref{sec:model}, we analyze the Poynting flux of the simulation data
in the cylindrical coordinates, $(R,\phi,z)$.
The vertical ($z$) component of Poynting flux in the MHD approximation
(see also Equation \ref{eq:totengMHDs}) can be reduced to 
\begin{equation}
  -\left[\frac{1}{4\pi}(\mbf{v}\times\mbf{B})\times\mbf{B}\right]_z = \frac{1}{4\pi}
  \left(v_z \mbf{B}_{\perp}^2 - B_z \mbf{v}_{\perp}\cdot\mbf{B}_{\perp}\right), 
\end{equation}
where the subscript $\perp$ indicates the two perpendicular ($R$ and
$\phi$) components with respect to the $z$ direction;
the first term corresponds to the Poynting flux carried by advected magnetic
energy and the second term denotes the Poynting flux by magnetic tension force.
If we examine this second term, we can estimate the net upgoing energy flux
of \Alfvenic perturbations.

In order to compare the energetics argument shown in Equation (\ref{eq:lumA}),
we take the areal integration of the \Alfvenic Poynting flux carried by MHD plasma along the $z$ direction\footnote{Here
  we assume the background is static.
  A general expression for the energy flux of \Alfvenic waves can be found
  in Equation (\ref{eq:Alflux}) in Appendix.},
\begin{equation}
  F_{\rm A} = -\frac{1}{4\pi}B_z \mbf{v}_{\perp}\cdot\mbf{B}_{\perp},
  \label{eq:FA}
\end{equation}
to derive ``\Alfvenic luminosity'',
\begin{equation}
  L_{\rm A}(z/R) = \int_{0}^{2\pi} d\phi \int_{R_{\rm in}}^{R_{\rm out}}RdR \left[
  F_{\rm A}\right]_{z/R}, 
  \label{eq:LA}
\end{equation}
where $F_{\rm A}$ is positive ($\mbf{v}_{\perp}\cdot\mbf{B}_{\perp}<0$)
  when flux is directed to $+B_z$, and we adopt $R_{\rm in}=0.23$ kpc and $R_{\rm out}=1.1$ kpc to cover the bulge
region. We do not take the inner boundary $=0.01$ kpc of the
simulation for $R_{\rm in}$ because we would like to avoid unphysical effects
of the boundary. For the radial integration, we take Poynting flux at the same
aspect ratio, $z/R$, instead of the same height, $z$, because the simulation, 
which was carried out in spherical coordinates, do not have data in the region
with large $z$ at small $R$. 

Figure \ref{fig:LA} shows the vertical profile of $L_{\rm A}$.
We note that the vertical boundaries of the 3D simulation by \citet{suz15} are located at $z/R=1.4$, and we do not consider the effect of the boundaries is sever in the plotting region, $|z/R|\le 1$.
One may recognize a characteristic feature in the plot: 
The direction of $L_{\rm A}$ is toward
the Galactic plane in $|z/R|\lesssim 0.6$; on the contrary it is directed
upward ($+$ for $z/R>0$ and $-$ for $z/R<0$) in $|z/R|\gtrsim 0.6$.
This indicates that \Alfvenic
Poynting flux is injected from the regions at $|z/H|\approx 0.6$ toward both
midplane and upper directions. This behavior is already observed in local
shearing box \citep{si09} and global \citep{si14} simulations, and
these regions are called injection regions.
While the magnetic energy is dominated by the thermal energy
(plasma $\beta \gg 1$) below the injection regions ($|z/R|\lesssim 0.6$),
the magnetic energy is comparable to the thermal energy ($\beta \approx 1$)
above the injection regions ($z/R|\gtrsim 0.6$). As a result, large-scale
channel-mode flows are developed by the MRI near the injection regions,
and breakups of these channel flows drive flows and Poynting fluxes to the
lower and upper directions. 

\begin{figure}
  \begin{center}
    \includegraphics[width=0.46\textwidth]{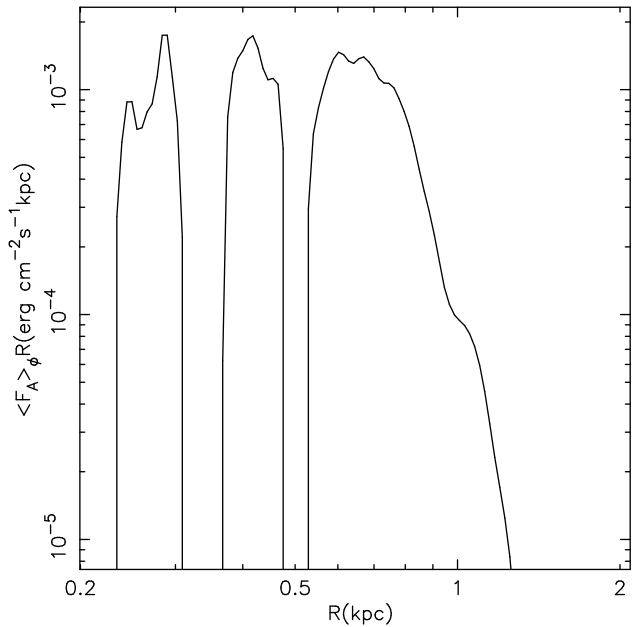}
  \end{center}
  \caption{Radial dependence of \Alfvenic energy flux multiplied by $R$ kpc
    averaged over time, azimuthal angle, and $0.75<|z/R|<0.9$. 
  \label{fig:FArdep}} 
\end{figure}

The time-averaged $L_{\rm A}$ during a period of 7.4 Myr
(solid line in Figure \ref{fig:LA}) reaches $\approx 4\times 10^{40}$erg s$^{-1}$
at $|z/H|\approx 0.8$, which is comparable to the estimate in Equation
(\ref{eq:lumA}). $L_{\rm A}$ at two snapshots (dotted and dashed lines) show
that $L_{\rm A}$ fluctuates with time and sometimes becomes
$\sim 10^{41}$erg s$^{-1}$. 

In order to examine spacial fluctuations of \Alfvenic flux, we plot the energy
flux averaged during 7.4 Myr over the full $2\pi$ azimuthal angle, 
\begin{equation}
  \langle F_{\rm A} \rangle_{\phi} = \int_{0}^{2\pi} d\phi F_{\rm A} / 2\pi, 
\end{equation}
at different rings, $R=0.28$, 0.58, and 0.78 kpc in Figure \ref{fig:FA}.
On the right axis, we show $\langle F_{\rm A}\rangle_{\phi}$ multiplied
by the bulge area $\pi (R_{\rm out}^2-R_{\rm in}^2)$ to give the unit of
luminosity, which can be directly compared to Figure \ref{fig:LA}.
These three locations give the similar tread with the injection
regions around $|z/R|\approx 0.5-0.8$, which was shown in Figure \ref{fig:LA}.
On the other hand, the energy flux is larger at inner locations,  and
at $R=0.28$ kpc $\langle F_{\rm A} \rangle_{\phi} \pi (R_{\rm out}^2-R_{\rm in}^2)$
is nearly $4\times 10^{41}$erg s$^{-1}$, which is $\approx 10$ times larger
than the time-averaged $L_{\rm A}$ in Figure \ref{fig:LA}.  

Figure \ref{fig:FArdep} presents the radial dependence of \Alfvenic
Poynting flux. We take the average in a region of $0.75<|z/R|<0.9$, which
is located above the injection region and show
$\langle F_{\rm A}\rangle_{\phi}R$ erg cm$^{-2}$s$^{-1}$kpc to see the
contribution to luminosity, Equation (\ref{eq:LA}).
This plot shows that the \Alfvenic Poynting flux equally contributes to
the luminosity in most of the bulge region, $R_{\rm in}<R<R_{\rm out}$
except for a few regions ($R\approx 0.25$ and 0.5 kpc) that temporally give
negative $F_{\rm A}$.

\section{Model for Alfv\'{e}nic wave-driven outflows}
\label{sec:model}
We consider the generation of outflows driven by low-frequency \Alfvenic waves
in a 1D flux tube that extends to the vertical direction,
$z$. We assume the cross section of
the flux tube expands with $z$ in the following manner
\citep{kh76,bre91,eve08}: 
\begin{equation}
  A = A_{\rm mid}\left[1+\left(\frac{z}{z_{\rm break}}\right)^2\right], 
  \label{eq:crosssection}
\end{equation}
where $A = A_{\rm mid}$ at the Galactic midplane, $z=0$. 
In this functional form the cross section is almost constant
in $z<z_{\rm break}$, and after that it expands with $\propto z^2$ in a spherical
manner at a large distance, $z > z_{\rm break}$ in the Galactic halo.

\begin{figure}
  \begin{center}
    \includegraphics[width=0.46\textwidth]{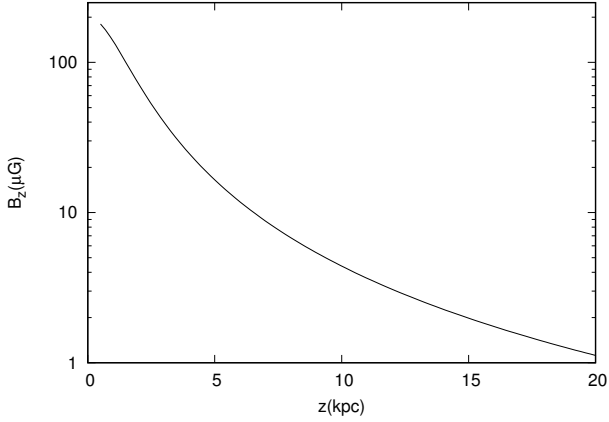}
  \end{center}
  \caption{Vertical profile of $B_z$.
  \label{fig:Bz}}
\end{figure}

We consider a 1D magnetic flux tube located at $R=R_0=0.3$ kpc, which is
slightly offset from the GC, whereas the choice of $R_0$ does not affect
our results unless we choose a large $R_0>1$ kpc.
The magnetic flux tube, which is fixed with time, is set up to follow
Equation (\ref{eq:crosssection}). The conservation of magnetic flux,
$\mbf{\nabla\cdot B=0}$ gives
\begin{equation}
  B_z = B_{\rm mid} A_{\rm mid} / A, 
  \label{eq:Bconserve}
\end{equation}
We assume $B_{\rm mid}=200$ $\mu$G at $z=0$ and $z_{\rm break}=1.5$ kpc.
These choices give the profile of $B_z$ as shown in Figure \ref{fig:Bz},
which is consistent with the observationally constrained magnetic field
strength, 5-20 $\mu$G, in the Fermi bubbles \citep{ack14}. 

We follow the time evolution of mass density, $\rho$, vertical velocity,
$v_z$, and internal energy of gas, $e_{\rm g}$, in this geometry of flux tubes
by our 2nd order Godunov method \citep{san99}.
The equation of mass conservation in the Lagrangian form is 
\begin{equation}
  \frac{d\rho}{dt} + \frac{\rho}{A}\frac{\partial}{\partial z}(Av_z) = 0. 
  \label{eq:cont}
\end{equation}
The momentum equation is 
\begin{equation}
  \rho \frac{dv_z}{dt} = -\frac{\partial}{\partial z}(p_{\rm g} + p_{\rm A})
  - \rho \frac{\partial\Phi}{\partial z}= 0, 
  \label{eq:eqm}
\end{equation}
where $p_{\rm g}$ is gas pressure, $p_{\rm A}$ is pressure of \Alfvenic waves,
which is modeled below, and $\Phi$ is a gravitational potential. 
The internal energy of gas is related to gas pressure and temperature via
\begin{equation}
  e_{\rm g} = \frac{1}{\gamma -1}\frac{p_{\rm g}}{\rho}
  = \frac{1}{\gamma -1}\frac{k_{\rm B}T}{\mu m_{\rm p}}, 
\end{equation}
where the ratio of specific heats is assumed to be $\gamma = 5/3$. 
The energy equation that determines $e_{\rm g}$ is
\begin{equation}
  \rho \frac{de_{\rm g}}{dt} + \frac{p_{\rm g}}{A}
  \frac{\partial}{\partial z}(Av_z) = H_{\rm A}
  \label{eq:gaseng}
\end{equation}
where $H_{\rm A}$ indicates the heating by the dissipation of
\Alfvenic waves, which is modeled below.
In Equation (\ref{eq:gaseng}), we do not take into account radiative cooling,
because its contribution is not substantial \citep{eve08}.

We handle the propagation and dissipation of \Alfvenic waves by the WKB
approximation \citep[][]{bel71,ac71,lc99}; instead of directly solving MHD wave
equations, we follow the variation of wave amplitude as a function of
$t$ and $z$.
The amplitudes of velocity and magnetic fields of \Alfvenic perturbations
are related via
\begin{equation}
  \mbf{v}_{\perp} = -\frac{\mbf{B}_{\perp}}{\sqrt{4\pi\rho}}.  
\end{equation}
Based on this relation, we define energy per mass, $e_{\rm A}$:
\begin{equation}
  e_{\rm A} \equiv \frac{1}{2}\mbf{v}_{\perp}^2 + \frac{\mbf{B}_{\perp}^2}{8\pi\rho}
  = \mbf{v}_{\perp}^2 = \frac{\mbf{B}_{\perp}^2}{4\pi\rho}. 
  \label{eq:eAdef}
\end{equation}
We consider \Alfvenic waves in a frequency range from $\omega_0$ to $\omega_1$,
\begin{equation}
  e_{\rm A} = \int_{\omega_0}^{\omega_1}d\omega \tilde{e}_{\rm A}(\omega)d\omega
  =\int_{\omega_0}^{\omega_1}d\omega \tilde{e}_{\rm A}(\omega_0)f(\omega/\omega_0)
  d\omega,
  \label{eq:wvengindeg}
\end{equation}
where we assume a power-law-dependence on frequency, $\omega$,
\begin{equation}
  f(\omega/\omega_0) = \left(\frac{\omega}{\omega_0}\right)^{-\alpha}.
  \label{eq:fomega}
\end{equation}
We do not assume a wide-band spectrum but a rather narrow-band spectrum
within a range of an order of magnitude between $\omega_0$ and $\omega_1$,
which is expected from magnetic activity with the characteristic timescale
$\sim 0.1-1$ Myr (see below for the specific choices of $\omega_0$,
$\omega_1$, and $\alpha$).  

Assuming \Alfvenic waves that propagate along the $+B_z$ direction and
the wavelengths are shorter than a typical scale, e.g. a pressure scale height
or a variation scale of \Alfven speed, of the background physical properties,
we can derive an equation
that describes the variation of $\tilde{e}_{\rm A}$ (see Appendix
for the detailed derivation): 
\begin{equation}
  \hspace{-0.5cm}
  \rho \frac{d\tilde{e}_{\rm A}}{dt} + \frac{1}{A}\frac{\partial}{\partial z}
  \left[A \rho \tilde{e}_{\rm A}\left(v_{\rm A} + \frac{1}{2}v_z\right)\right]
  - \frac{v_z}{2}\frac{\partial}{\partial z}(\rho \tilde{e}_{\rm A})
  = - \gamma_{\rm A}\rho \tilde{e}_{\rm A}, 
  \label{eq:alfeng}
\end{equation}
where $v_{\rm A}=B_z/\sqrt{4\pi\rho}$, $\gamma_{\rm A}$ is the damping rate
of \Alfvenic waves, and 
we neglect nonlinear interactions between different waves and assume
that waves with different frequencies evolve independently.
In this case, we can derive
an equation that describes the evolution of the spectral index (see Appendix):
\begin{equation}
  \frac{d\alpha}{dt} + v_{\rm A}\frac{\partial \alpha}{\partial z}
  = \frac{\gamma_{\rm A}(\omega) - \gamma_{\rm A}(\omega_0)}{\ln(\omega/\omega_0)}.
  \label{eq:alfidx}
\end{equation}
For the wave component, we solve Equation (\ref{eq:alfeng}) for $\omega
=\omega_0$ and Equation (\ref{eq:alfidx}), and then, we can derive the
energy density, $e_{\rm A}$, integrated over $\omega$ from  Equation
(\ref{eq:wvengindeg}).
The heating rate by the dissipation of \Alfvenic waves (Equation
\ref{eq:gaseng}) can be derived from the frequency-integrated damping rate,
\begin{equation}
  H_{\rm A}=\langle \gamma_{\rm A}\rho e_{\rm A} \rangle
  = \int_{\omega_0}^{\omega_1}d\omega \gamma_{\rm A}(\omega)\rho
  \tilde{e}_{\rm A}(\omega).
\end{equation}

The gravitational potential, $\Phi$, in Equation (\ref{eq:eqm})
consists of four components,
\begin{equation}
  \Phi = \sum_{i=1}^{4}\Phi_i, 
\end{equation}
where each component corresponds to the SMBH
at the Galactic center ($i=1$), the bulge ($i=2$), the disk ($i=3$),
and the dark halo ($i=4$).
For the SMBH, we adopt a point-mass, $M_1=4.4\times 10^6M_{\odot}$,
at the GC \citep{gen10}, where $M_{\odot}$ is the solar mass. 
For the bulge and disk components, we use a gravitational potential introduced
by \citet{mn75}: 
\begin{equation}
\Phi_{i=2,3}(R,z)= \frac{-GM_i}{\sqrt{R^2+(a_i+\sqrt{b_i^2+z^2})^2}},
\label{eq:potential}
\end{equation}
where 
$M_2 = 2.05\times 10^{10}M_{\odot}$,
$a_2 = 0$, $b_2 = 0.495$ kpc, $M_3 = 25.47\times 10^{10}M_{\odot}$,
$a_3 = 7.258$ kpc, and $b_3 = 0.52$ kpc.
For the dark halo, we adopt the NFW density profile \citep{nav96}, which gives
the following form of the  gravitational potential \citep[e.g.,][]{kuz09}, 
\begin{equation}
  \Phi_4 = -4\pi G \rho_{\rm h,0} r_{\rm h}^3\frac{1+r/r_{\rm h}}{r}, 
\end{equation}
where $r = \sqrt{R^2 + z^2}$ is the spherical radius,  and we assume
$r_{\rm h} = 10.7$ kpc and $\rho_{\rm h,0}=1.82\times 10^{-2}M_{\odot}$pc$^{-3}$
\citep{sof15}.

\Alfvenic perturbations injected by the magnetic activity in the Galactic bulge
region travel upward into the Galactic halo.
We focus on low-frequency \Alfvenic waves with the typical wavelength $0.1-1$
kpc. Such long-wavelength waves are not subject to nonlinear Landau damping,
which is effective for waves with the wavelength of an order of the gyration
scale \citep[][see also Section \ref{sec:intro}]{kp69,bl08}.  
We also restrict our focus on linear \Alfven waves, of which the amplitude is
significantly smaller than the local \Alfven speed.
Such linear waves in the MHD regime do not suffer dissipation in homogeneous
media. However, if the background magnetic medium is turbulent, \Alfven waves
are damped because of the meandering magnetic field \citep{laz16}.
Indeed, various sources of turbulence in the Fermi bubbles have been proposed,
such as Kelvin-Helmholtz, Rayleigh-Taylor, or Richtmyer-Meshkov instabilities
behind shock fronts \citep{gm12,ino12,sas15}, and infalling stars and gas clouds
to the central SMBH \citep{che15}.

We consider MHD turbulence injected from large-scale sources with size
$\sim 1$ kpc. The linear damping rate of \Alfven waves can be found in
\citet{laz16}: 
\begin{equation}
  \gamma_{\rm A} = \frac{v_{\rm A} M_{\rm A}^{4/3}\sin^{2/3} \theta}{L^{1/3}\lambda^{2/3}}
  \approx \frac{v_{\rm A}M_{\rm A}^2}{L^{1/3}\lambda^{2/3}}
  = \frac{v_{\rm A}M_{\rm A}^2\omega^{2/3}}{L^{1/3}(v_{\rm A} + v_z)^{2/3}},
  \label{eq:damping}
\end{equation}
where $L$ is an injection scale of turbulence,
\begin{equation}
  \lambda = (v_{\rm A} + v_z) \tau_{\rm w} = (v_{\rm A} + v_z)
  \left(\frac{2\pi}{\omega}\right)
  \label{eq:wavelengthdef}
\end{equation}
is the wavelength of propagating waves, $M_{\rm A}$ is the Mach number of
turbulence, and $\theta$ is the angle between the direction of the wave
propagation and the background magnetic field.
We focus on the \Alfven waves that propagate in parallel with
the average direction of the background field; in this case the angle $\theta$
can be approximated by the meandering angle of magnetic field lines,
\begin{equation}
  \sin \theta \approx \delta B/B_z = M_{\rm A},
  \label{eq:angleapprox}
\end{equation}
where $\delta B$ is the amplitude of magnetic turbulence. 
When deriving the final approximate expression of Equation (\ref{eq:damping}),
we used Equation (\ref{eq:angleapprox}). 

We consider large-scale sources for the turbulence as discussed above,
and assume $L=1$ kpc in this paper.
Wave sources are various magnetic activities in the bulge regions, such as
breakups of channel flows, buoyantly rising magnetic loops
\citep{fuk06,tor10a,tor10b},  and spring-like helical magnetic
structure \citep{eno14}.
For rough estimates, let us consider magnetic structure with field strength of 
$B\sim 300$ $\mu$G and a size of $l_{\rm B}\sim 0.1$ kpc, which generates
\Alfvenic Poynting flux. Then, we get a characteristic period, 
\begin{eqnarray}
  \tau_{\rm w} &\sim& l_{\rm B} / v_{\rm A} \nonumber \\
  &\approx& 0.3\;{\rm Myr} \left(\frac{l_{\rm B}}{0.1\;{\rm kpc}}\right)
  \left(\frac{B}{300\;\mu{\rm G}}\right)^{-1}
  \left(\frac{n}{10\;{\rm cm}^{-3}}\right)^{1/2}. 
\end{eqnarray}
Following the above estimate, we consider waves in a range between
$\tau_{\rm w,1} = 0.1$ Myr and $\tau_{\rm w,0} = 1$ Myr, and inject \Alfvenic
waves in $\omega_0 < \omega < \omega_1$ with a power-law index, $\alpha=1$,
where $\omega_0 = 2\pi/\tau_{\rm w,0}$ and $\omega_1 = 2\pi/\tau_{\rm w,1}$.
Note that $\alpha=1$ indicates that waves with different $\omega$ 
are equally injected in logarithmic spacing, $d\log \omega$. 
We leave the uncertainty of the wave damping to $M_{\rm A}$; choosing
a smaller $L$ 
is practically the same as setting a larger
$M_{\rm A}$. We analyze the effect of the wave damping by changing $M_{\rm A}$.  

For later analyses, we would like to estimate a typical wavelength:
\begin{equation}
  \lambda \sim v_{\rm A} \tau_{\rm w} \approx 0.3\;{\rm kpc}
  \left(\frac{v_{\rm A}}{1000\;{\rm km\; s^{-1}}}\right)
  \left(\frac{\tau_{\rm w}}{0.3\;{\rm Myr}}\right), 
  \label{eq:lambdaest}
\end{equation}
where we neglect $v_z$ in Equation (\ref{eq:wavelengthdef}) and adopt
a standard value for the \Alfven speed in the halo region. 


\begin{figure*}
  \begin{center}
    \includegraphics[width=0.9\textwidth]{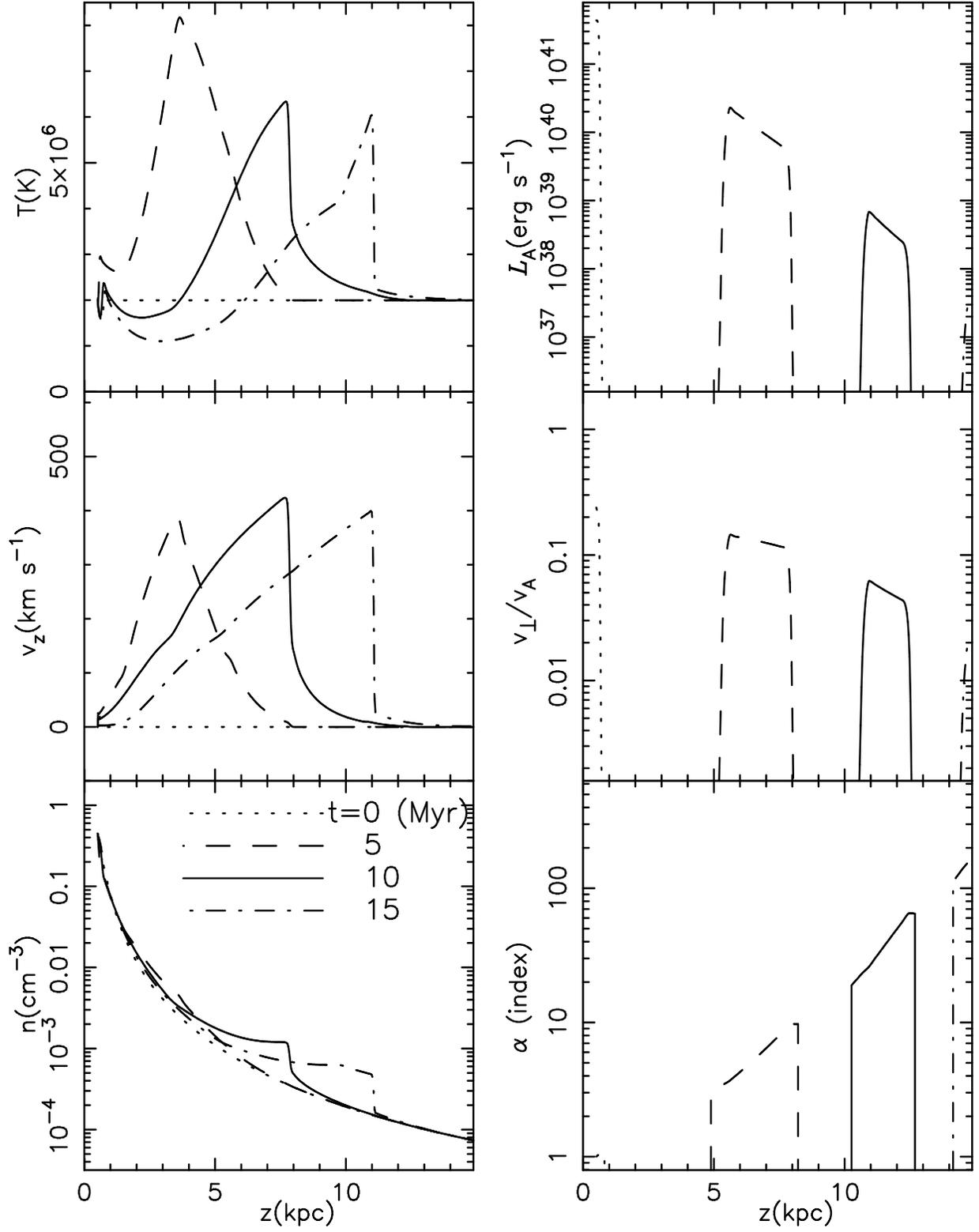}
  \end{center}
  \caption{Time evolution of vertical distributions of physical quantities
    for the case with turbulent Mach number $M_{\rm A}=0.5$ and injection time
    $\tau_{\rm inj} = 2$ Myr.
    Dotted, dashed, solid, and dot-dashed lines indicate $t=0$, 5, 10,
    and 15 Myr, respectively. 
    {\it Top Left}: Temperature, $T$ (K).
    {\it Middle Left}: Vertical velocity, $v_z$ (km s$^{-1}$).
    {\it Bottom Left} Particle number density, $n$ (cm$^{-3}$).
    {\it Top Right}: \Alfvenic luminosity, ${\cal L}_{\rm A}$ (erg s$^{-1}$), 
    described by Equation (\ref{eq:calL}). 
    {\it Middle Right}: Nonlinearity of the \Alfvenic perturbations,
    $v_{\perp}/v_{\rm A}$.
    {\it Bottom Right}: Spectral index, $\alpha$, of the \Alfvenic
    perturbations. 
    {\it Movie} is also available as a supplementary file and
    at http://ea.c.u-tokyo.ac.jp/astro/Members/stakeru/research/movie/index.html.
    \label{fig:tev05} }
\end{figure*}



\begin{figure}
  \begin{center}
    \includegraphics[width=0.45\textwidth]{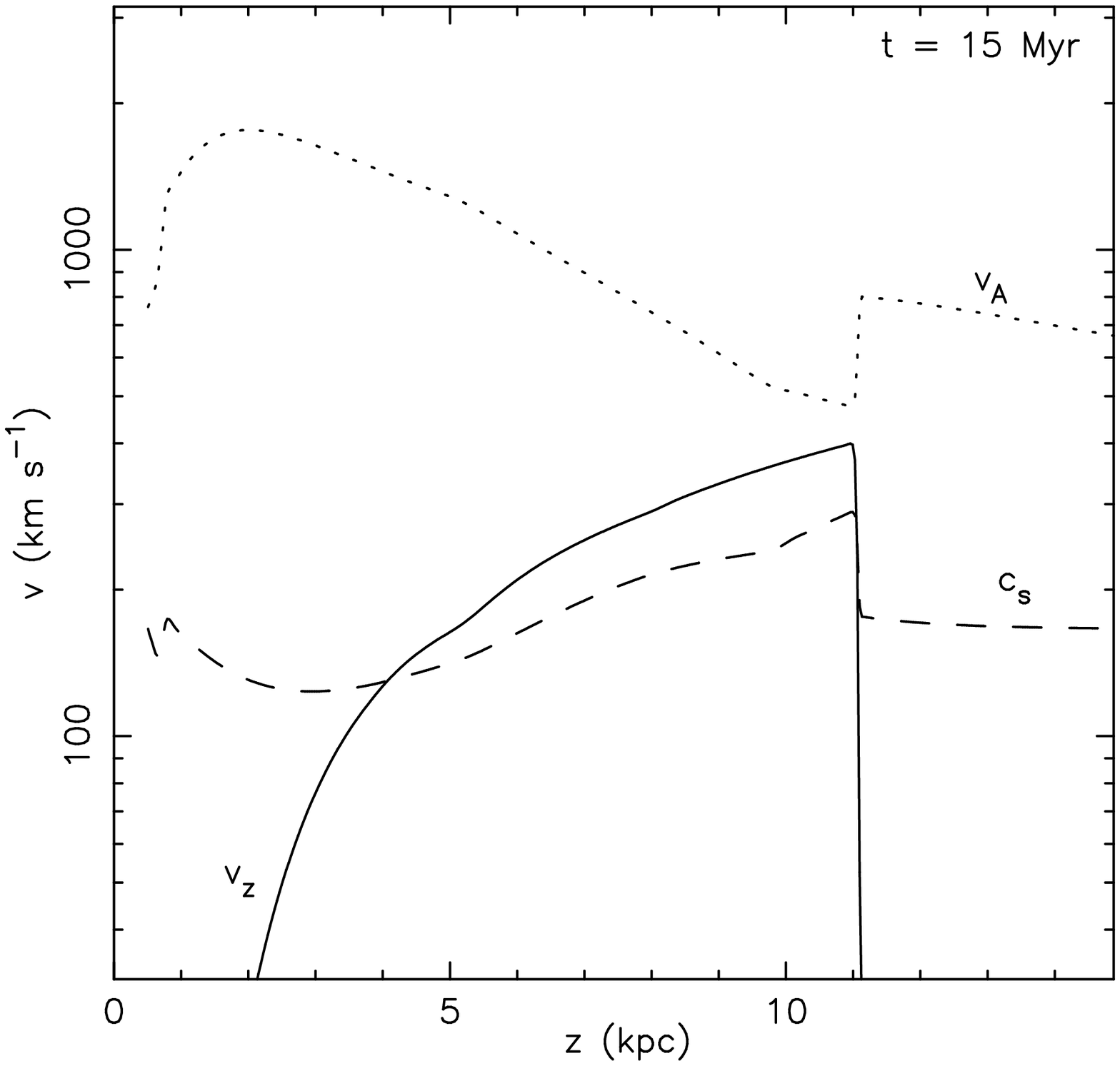}    
  \end{center}
  \caption{Comparison of vertical velocity, $v_z$, (solid), \Alfven speed,
    $v_{\rm A}$, (dotted), and sound speed, $c_{\rm s}$, (dashed) at $t=15$ Myr.
  \label{fig:vstr}}
\end{figure}

The calculation region covers from the bottom boundary at $z=0.5$ kpc to
the top boundary $= 23$ kpc.
Initially we set up the hydrostatic equilibrium with a constant
temperature, $T_{\rm halo}=2\times 10^6$ K, in the Galactic halo \citep{mb16}.
At the bottom boundary at $z=0.5$ kpc, we set the particle number density
$n=4.5\times 10^{-1}$ cm$^{-3}$, which corresponds to the electron number
density $\approx 2.4\times 10^{-1}$ cm$^{-3}$ for $\mu=0.6$.
The hydrostatic density structure gives $n=1.9\times 10^{-4}$cm$^{-3}$
at $z=10$ kpc.
We note that the density shown in Figure \ref{fig:snp3D} is $n\approx 10$ cm$^{-3}$ at $z=0.5$ kpc, which is larger than the density adopted at the bottom boundary of the present setup.
  This is because of the difference of the adopted temperatures.
  In the 3D simulation of Figure \ref{fig:snp3D}, the temperature is assumed to be locally isothermal with $(2-5)\times 10^5$K in the bulge region, which reflects the velocity dispersion of gas clouds.
This temperature is lower than the temperature at the bottom boundary of the present model, and therefore, we adopt the lower density to give similar gas pressure ($\propto nT$). 

We inject \Alfvenic perturbations by giving non-zero $e_{\rm A,0}$ from
the bottom boundary, while the other variables, $T$, $v_z$, and $n$, are fixed
at the bottom boundary.
At the outer boundary, we prescribe the outgoing condition
by using equations for characteristics \citep{si05,si06}.
We update $\rho$, $v_z$, $T$ (or $e_{\rm g}$), and
$e_{\rm A}$ by solving Equations (\ref{eq:cont}) -- (\ref{eq:gaseng}) and
Equation (\ref{eq:alfeng}) during $\tau_{\rm sim}=20$ Myr.
We test two types of the energy injection: One is a continuous injection,
in which $e_{\rm A,0}$ is constant throughout $\tau_{\rm sim}$.
The other is a temporal injection, in which we input $e_{\rm A,0}\ne 0$ during 
the initial $t<\tau_{\rm inj}$ and switch off $e_{\rm A,0}$ at $t=\tau_{\rm inj}$
to keep $e_{\rm A,0}=0$ during $\tau_{\rm inj}<t<\tau_{\rm sim}$. 
The total injected energy is determined from the time-integrated $L_{\rm A}$
given by the 3D MHD simulation introduced in Section \ref{sec:3DMHD},   
\begin{equation}
  \rho v_{\rm A} e_{\rm A,0} \pi (R_{\rm out}^2 - R_{\rm in}^2)\tau_{\rm inj}
  = L_{\rm A} \tau_{\rm sim}, 
  \label{eq:wvinj0}
\end{equation}
where we adopt $L_{\rm A}=4.3\times 10^{40}$erg s$^{-1}$ (Figure \ref{fig:LA}).
Table \ref{tab:waveinj} summarizes $e_{\rm A,0}$ that satisfies Equation
(\ref{eq:wvinj0}). Here, the velocity amplitude, $v_{\perp,0}$, is derived
from Equation (\ref{eq:eAdef}). 

\begin{table}
  \begin{tabular}{|c|c|c|c|}
    \hline
    Injection Type & $\tau_{\rm inj}$ (Myr) & $e_{\rm A,0}$ (erg g$^{-1}$) & $v_{\perp,0}$ (km s$^{-1}$)  \\
    \hline
    \hline
    Continuous & 20 & $3.3\times 10^{13}$ & 57 \\
    \hline
    Temporal & 2 & $3.3\times 10^{14}$ & 180 \\
    \hline
  \end{tabular}
  \caption{Parameters for the injected wave energy. $v_{\perp,0}$ is derived
    from $e_{\rm A,0}$ by Equation (\ref{eq:eAdef}).
    \label{tab:waveinj}}
\end{table}

\section{Results}
\label{sec:res}
\begin{figure*}
  \begin{center}
        \includegraphics[width=0.9\textwidth]{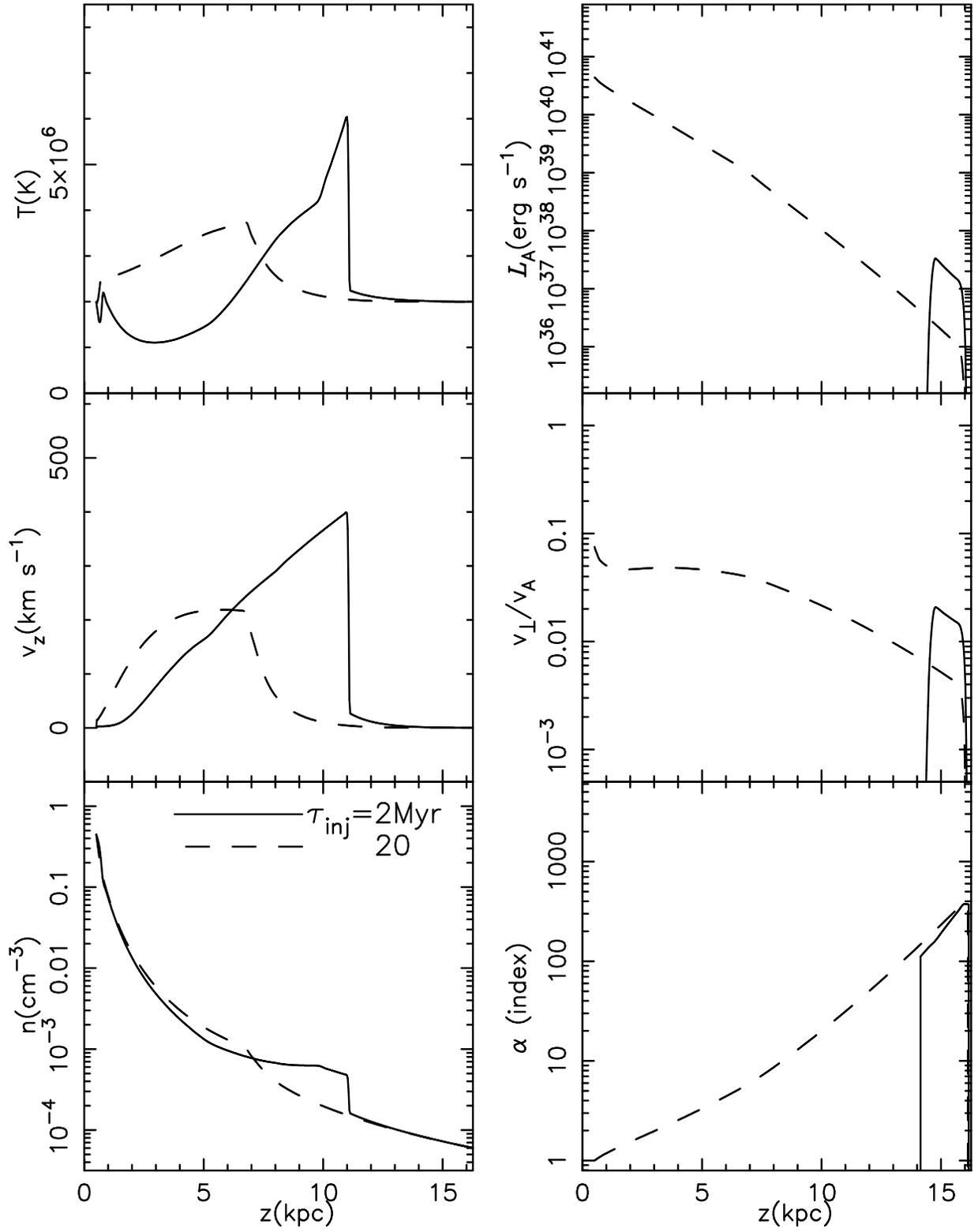}
  \end{center}
  \caption{Comparison of cases with different $\tau_{\rm inj}=2$ Myr (solid)
    and 20 Myr (dashed) for $M_{\rm A}=0.5$ at $t=15$ Myr.
    The six panels are the same as in Figure \ref{fig:tev05}.
  \label{fig:tinjdep}}
\end{figure*}

\begin{figure*}
  \begin{center}
    \includegraphics[width=0.9\textwidth]{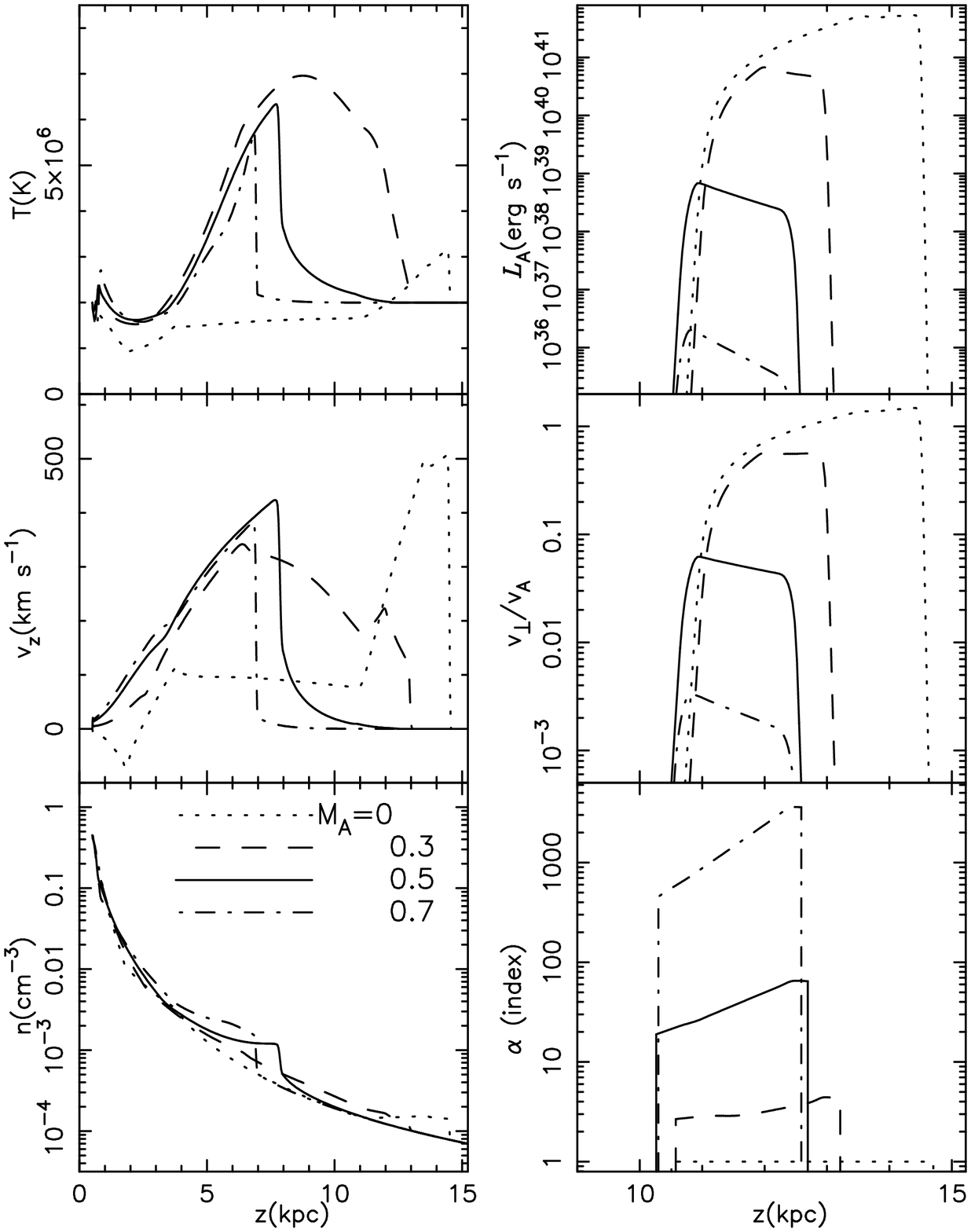}
  \end{center}
  \caption{Comparison of cases with different $M_{\rm A}=0$ (dotted),
    0.3 (dashed), 0.5 (solid), and 0.7 (dot-dashed) for $\tau_{\rm inj}=2$ Myr
    at  $t=10$ Myr. The six panels are the same as in Figure \ref{fig:tev05}.
  \label{fig:Madep}}
\end{figure*}

Figures \ref{fig:tev05} presents how an outflow is driven
by the injected \Alfvenic Poynting flux in the case with $M_{\rm A}=0.5$ and
$\tau_{\rm inj}=2$ Myr. 
We show the evolution of physical quantities of the gas in the left panels and
physical properties of the \Alfvenic waves in the right panels. 
The 
top right panel shows the quantity, 
\begin{equation}
  {\cal L}_{\rm A}\equiv S_{\rm A}A \pi (R_{\rm out}^2 - R_{\rm in}^2), 
  \label{eq:calL}
\end{equation}
in units of luminosity (erg s$^{-1}$) for the comparison to Figures \ref{fig:LA}
\& \ref{fig:FA} in Section \ref{sec:3DMHD}, 
where
\begin{equation}
  S_{\rm A} = \rho e_{\rm A} \frac{(v_{\rm A}+v_z)^2}{v_{\rm A}} 
\end{equation}
is an adiabatic constant called wave action
\citep[][see also Equation \ref{eq:waveaction} in Appendix]{jac77}.
We call ${\cal L}_{\rm A}$ ``wave action luminosity'' in this paper.

${\cal L}_{\rm A}$ decreases rapidly with $z$, 
and the nonlinearity, 
$v_{\perp}/v_{\rm A}$, (middle right panel) is kept $\lesssim 0.2$, which
justifies the linear damping process adopted here as a dominant mechanism.
Because of the rapid damping of the Poynting flux, the gas is
effectively heated up (top left panel). 
One can clearly
see the formation of a shock front that moves upward. The temperature at the
shock front reaches $T=8\times 10^6$ K at $t=5$ Myr and $T=6\times 10^6$ k
at $t=10$ Myr. Accordingly, the gas is accelerated to $v_z = 400-500$
km s$^{-1}$ (middle left panel).
The shock heating also contributes to the mass loading to the upper
layer, and the density increases to $n\approx
(6-7)\times 10^{-4}$ cm$^{-3}$ at $z=10$ kpc
when the shock front passes there at $t=13.5$ Myr. These physical quantities
are consistent with observationally derived values, $T=(4-5)\times 10^6$ K and
$n\approx 10^{-3}$cm$^{-3}$ introduced in Section \ref{sec:intro} \citep{mb16}.

Comparing the top right panel to the left three panels in Figure
\ref{fig:tev05}, we may find that the front of the \Alfvenic wave travels
faster than the shock front because the \Alfven speed is higher than the sound
speed, which is shown in Figure \ref{fig:vstr}. On the other hand,
we cannot find any signature in the hydrodynamical quantities, $T$, $v_z$,
and $n$, at the front of the \Alfvenic wave.
In this case with $M_{\rm A}=0.5$, the Poynting flux is damped at
a lower altitude. Therefore, they cannot significantly
contribute to the acceleration by the magnetic pressure with
\Alfvenic perturbations (Equation \ref{eq:eqm}) at a higher altitude $\gtrsim
5$ kpc, but simply heats up the gas by the dissipation of the Poynting flux
below $z < 5$ kpc.

The bottom right panel of Figure \ref{fig:tev05} shows that the spectral index,
$\alpha$, increases as the \Alfvenic waves propagate. This is because waves
with higher $\omega$ are damped more rapidly (see Equation \ref{eq:damping})

Figure \ref{fig:vstr} shows that the outflow is supersonic ($v_z>c_{\rm s}$)
but slightly sub-\Alfvenic even at the shock front. However, $v_z$ exceeds
the local escape velocity $\sim 100$ km s$^{-1}$ at $z\approx 10$ kpc,
and therefore, this outflow will not return back but further streams outward
and escapes from the gravity of the Galaxy. 

The \Alfven velocity, $v_{\rm A}$($\propto B_z/\sqrt{n}$),
increases rapidly in $z<2$ kpc owing to the
rapid drop of the density, $n$, there (see Figure \ref{fig:tev05}). However,
it gradually decreases because the effect of the decrease of $B_z$ by
the expansion of the magnetic flux tube
(Equations \ref{eq:crosssection} \& \ref{eq:Bconserve}) surpasses the effect
of the decrease of $n$. The mass supply from the lower region
additionally slows down the decrease of $n$ with $z$.
The profile of $v_{\rm A}$ is qualitatively similar to that obtained in
the solar atmosphere \citep[e.g.,][]{ver12}; the \Alfven velocity increases
with height in the corona because of the decrease of the density; however,
it gradually decreases with height in the solar wind because the decrease
of the density becomes gradual by the mass supply from the solar wind
in addition to the decrease of the magnetic field strength.

We should note that the detailed profile of $v_{\rm A}$ in Figure
\ref{fig:vstr} depends on the properties of the magnetic flux tube.
If we adopt smaller $B_{\rm mid}$
and/or smaller $z_{\rm break}$, $B_z$ and $v_{\rm A}$ at higher altitudes are
reduced.
In this case, the location of the wave dissipation is shifted to a lower
altitude, namely it is equivalent to enhanced wave damping.
In other words, a different choice of the magnetic flux tube is effectively
replaced by a different choice of $M_{\rm A}$ that controls wave dissipation
in our formulation. Therefore, we fix the parameters of the magnetic flux
tube and focus on the dependence on $M_{\rm A}$ below.

Figure \ref{fig:tinjdep} compares two cases with different $\tau_{\rm inj}$ for a
fixed $M_{\rm A}=0.5$.
Since the total energies injected from the lower boundary are the same in the
two cases, the input energy flux is smaller for the case with the longer
injection, $\tau_{\rm inj}=20$ Myr. 
This case also shows the formation of a shock wave, although its amplitude is
smaller than in the temporal injection case ($\tau_{\rm inj}=2$ Myr).
The peak temperature and velocity are also lower for longer $\tau_{\rm inj}$.

Figure \ref{fig:Madep} compares cases with different $M_{\rm A}$, which directly
controls the damping rate of the Poynting flux (Equation \ref{eq:damping}),
for a fixed $\tau_{\rm inj}=2$ Myr. 
In the no damping case ($M_{\rm A}=0$) the Poynting flux does not contribute to
the heating but transfer the only momentum to the gas; the heating of the gas
is possible only by adiabatic compression. Hence, the temperature does not
increase so much.
On the other hand, the gas is accelerated mainly by the magnetic pressure to
$> 500$ km s$^{-1}$

The outflow structure of the case with $M_{\rm A}=0.3$ shows intermediate
properties between the case with $M_{\rm A}=0.5$ and the case $M_{\rm A}=0$; 
both energy (heating) and momentum
(acceleration) transfers from the Poynting flux to the gas are important.
Comparison of the middle-left ($v_z$) panel to the right panels shows
that the outflow is first led by the front of the injected \Alfvenic flux.
This illustrates that the initial upflow is driven directly by the magnetic
pressure, which was not seen in the fast damping cases with $M_{\rm A}=0.5$
and $0.7$. One can see that the initial $v_z$ peak is followed
by 
the second peak, which corresponds to the shock front seen in
the fast damping cases, pushed by the gas pressure and travels with speed of
$\sim c_{\rm s}$.
Therefore, we can understand that the initial front is created by the direct
momentum transfer from the Poynting flux and the second peak is formed as a
result of the heating by the wave dissipation. 

The case with $M_{\rm A}=0.3$ also shows that the gas is heated up to
$T=(6-7)\times 10^6$ K, which is comparable to the
temperatures obtained in the case with $M_{\rm A}=0.5$. 
However, the density is not so high as that for $M_{\rm A}=0.5$, because
the heating near the footpoint is weaker and sufficient mass loading is not
achieved.
The middle right panel shows that the nonlinearity is kept $\approx 0.5$
even though the wave action luminosity decreases because of the wave damping.
This is mainly because the \Alfven speed decreases 
by the expansion of the flux tube. 
In such circumstances,
nonlinear damping processes (see Subsection \ref{sec:nonl}) would efficiently
work to suppress the nonlinearity.
Therefore, in realistic situations, the wave dissipation would be enhanced
to further heat up the gas.

\begin{figure}
  \begin{center}
    \includegraphics[width=0.5\textwidth]{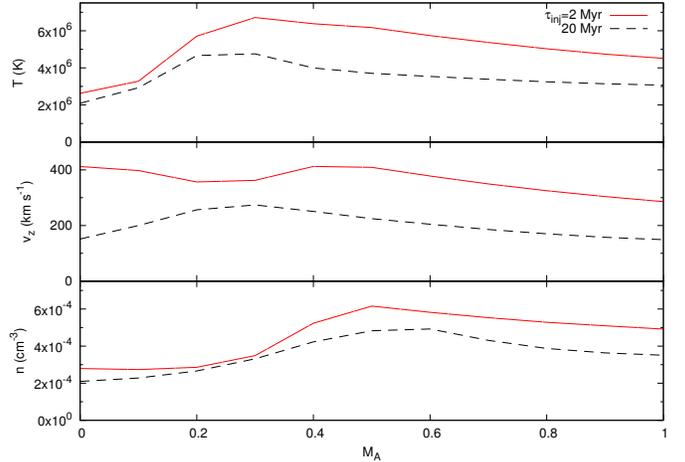}
  \end{center}
  \caption{Dependence of the maximum temperature (top), vertical velocity
    (middle), and density (bottom) at $z=10$ kpc on $M_{\rm A}$.
    Note that $M_{\rm A}$ determines the damping rate of the Poynting
    flux. Red solid and black dashed lines correspond to the cases for
    the temporal injection ($\tau_{\rm inj}=2$ Myr) and the cases for the
    continuous injection ($\tau_{\rm inj}=20$ Myr), respectively.
    \label{fig:peaks}
  }
\end{figure}

Figure \ref{fig:peaks} presents the maximum temperature, the maximum vertical
velocity, and the maximum density measured at $z=10$ kpc for cases with
different $M_{\rm A}$.
The top panel shows that the initial temperature ($=2\times 10^6$ K) is
kept for the no dissipation cases, $M_{\rm A}=0$.
The maximum temperature $\approx 7\times 10^6$ K for
$\tau_{\rm inj}=2$ Myr and $\approx 5\times 10^6$ K for $\tau_{\rm inj}=20$ Myr
is obtained for moderately small dissipation, $M_{\rm A}\approx 0.3$.
This is because the heating is kept up to higher locations $z \gtrsim 10$ kpc
on account of the moderately slow dissipation of the Poynting flux.

The middle panel shows that the temporal injection
($\tau_{\rm inj}=2$ Myr) cases give faster $v_z\approx 300-400$ km s$^{-1}$ than
the continuous injection ($\tau_{\rm inj}=20$ Myr) cases as expected.
The temporal cases shows a bimodal tread with two maximums of $v_z$ at
$M_{\rm A}=0$ and 0.4-0.5. The first peak at $M_{\rm A}=0$ is from the outflow
directly driven by the magnetic pressure of the \Alfvenic Poynting flux,
while the second peak at $M_{\rm A}=0.4-0.5$ is from the outflow driven
by the gas pressure (Figure \ref{fig:Madep}). 

The bottom panel shows a single maximum for $n$ at $M_{\rm A}\approx 0.5-0.6$
in both cases with $\tau_{\rm inj}=2$ and 20 Myr. This is qualitatively similar to
the temperature (top panel), but the peak location is shifted to larger
$M_{\rm A}$, because sufficient heating by the wave dissipation near the
footpoint is necessary to drive dense outflows \citep[e.g.,][]{lc99}.

\section{Discussion}
\label{sec:dis}
We have studied how outflows are driven by long-wavelength \Alfvenic waves
from magnetic activity in the bulge. 
In this section we discuss limitations regarding approximations we have assumed
when handling the propagation and dissipation of the \Alfvenic Poynting flux.

\subsection{Nonlinear Effects}
\label{sec:nonl}
We considered turbulent damping for the main dissipation channel of the
\Alfvenic waves. This mechanism is essentially a linear process, and therefore it
is dominant for waves for $v_{\perp}/v_{\rm A}\ll 1$.
Figure \ref{fig:Madep} shows that this condition is satisfied for the cases with
moderately large dissipation, $M_{\rm A}\ge 0.5$.
However, this is not the case for $M_{\rm A}\le
0.3$ because the amplitude becomes $v_{\perp}/v_{\rm A}\gtrsim 0.5$ so that the
\Alfven waves are nearly nonlinear.
For these waves, nonlinear damping processes probably operate to
enhance the dissipation.

For example, direct steepening of wave fronts occurs effectively.
As a result, \Alfvenic waves steepen to fast MHD shock trains
and finally dissipate \citep{suz04}. These are small-scale fast MHD
shocklets and different from the large-scale hydrodynamical shock wave
excited by the initial dissipation of the injected \Alfvenic Poynting flux
presented in Section \ref{sec:res} and Figures \ref{fig:tev05} \&
\ref{fig:Madep}. However, they are also expected to contribute to the heating
and driving hot outflows.
In addition, longitudinal slow MHD waves are also excited by the fluctuating
magnetic pressure of the \Alfvenic waves \citep{ms12,ms14,miy14}.
These small-scale slow MHD ($\approx$ acoustic) shocklets further contribute
to the heating. 
This nonlinear mode conversion mechanism is also important in terms of
the generation of density perturbation, which plays a role in the reflection
of \Alfvenic waves \citep{si06} as shown below.

Nonlinear interactions between different waves will be also efficient. 
In particular, the interaction between counter-propagating \Alfven waves as
a result of reflection (see Section \ref{sec:reflection}) excites MHD turbulence
\citep[e.g.,][]{gs95,cl03}, which increases $M_{\rm A}$.
Although we have assumed that the \Alfven waves propagate in the turbulent
media with a constant $M_{\rm A}$, this argument shows that in reality
propagating waves probably affect properties of the turbulence.
Therefore, the wave amplitude and $M_{\rm A}$ of the background turbulence
are regulated each other, which is to be pursued in future works.

\subsection{Reflection}
\label{sec:reflection}
We treated the propagation of the \Alfvenic waves, instead of directly solving
wave equations, by updating energy $e_{\rm A}$ and the spectral index
$\alpha$ with time under the WKB approximation, which is satisfied under
the condition that
the wavelength is sufficiently smaller than the variation scale of the \Alfven
speed. By this treatment, we could avoid numerical diffusion that
  particularly affects propagation of high-frequency (short-wavelength) waves.
On the other hand, unless this condition is satisfied, \Alfven waves are
reflected by the change of the wave shape \citep{si06,sy16}.
We adopted the wide-band spectrum (Equation \ref{eq:wvengindeg}).
The lower frequency boundary, $\omega_0=1/1$ Myr corresponds to the longest
wavelength we consider, which is an order of 1 kpc
(Equation \ref{eq:lambdaest}). 
The typical variation scale of $v_{\rm A}$ is comparable to 1 kpc
in $z\lesssim 2$ kpc,
and moderately larger than 1 kpc in $z\gtrsim 1$ kpc (Figure \ref{fig:vstr}).  
Therefore, the \Alfven waves near $\omega=\omega_0$ are expected to be subject
to reflection at a low altitude. The reflected waves eventually interact
with pre-existing upgoing waves to excite \Alfvenic turbulence as discussed
in Section \ref{sec:nonl}.

\subsection{Other Sources}
Although in this paper we only investigated the role of low-frequency \Alfvenic
Poynting flux with the period of 0.1 -- 1 Myr,
cosmic rays are believed to be a reliable candidate to drive large-scale
outflows \citep{bre91,eve08}; the effective pressure of the cosmic
rays pushes the halo gas upward, and high-frequency \Alfven
waves are excited via the streaming instability from the cosmic rays
\citep{wen68}. These waves are eventually damped by both
nonlinear Landau process \citep{kp69,bl08} and MHD turbulence
\citep{laz16}, which plays a role in the heating of the gas. 

These consecutive processes additionally contribute to accelerating
and heating up the halo gas.
In this sense, the results shown in the present paper give a lower
bound for the temperature and velocity of the outflows.

\section{Summary \& Conclusion}
We studied roles of low-frequency \Alfvenic waves with period of 0.1 -- 1
Myr excited by magnetic activity, such as buoyantly rising magnetic loops
and breakups of channel flows, in the Galactic bulge in driving large-scale
outflows.
We inspected the global 3D MHD simulation by \citet{suz15} and found that
the time-averaged \Alfvenic luminosity is $10^{40}-10^{41}$erg s$^{-1}$.
We input this level of the \Alfvenic Poynting flux from the footpoint of
a magnetic flux tube that expands into the Galactic halo.
We performed time-dependent hydrodynamic simulations with taking into account
the propagation and dissipation of \Alfvenic waves. 
We considered a linear dissipation mechanism of the \Alfvenic waves in
turbulent media developed by \citet{yl02,fg04,laz16}. 

Our model has essentially two parameters, the turbulent Mach number, $M_{\rm A}$,
that controls the dissipation of the \Alfvenic waves, and the duration of
the energy injection, $\tau_{\rm inj}$.
Our calculation shows that the basic thermal properties of the Fermi bubbles are
well explained by cases with nearly transonic turbulence, $M_{\rm A}\approx
0.5$, and an temporal injection, $\tau_{\rm inj} = 2$ Myr.
This result shows that the magnetic activity in the Galactic bulge can
potentially give a significant contribution to driving the Fermi-bubbles and
large-scale outflows, although our treatment for the \Alfvenic Poynting flux,
which is a simplified one, need further elaboration in future studies.

This work was supported in part by Grants-in-Aid for 
Scientific Research from the MEXT of Japan, 17H01105 (TKS), and
  the NSF grants DMS 1622353 (AL). T.K.S. thanks
  Prof. Yasuo Fukui for fruitful discussion.

\begin{appendix}
\section{Derivation of Wave Energy Equation and Wave Action}
A general MHD expression of the total energy conservation is 
\begin{equation}
  \frac{\partial}{\partial t}\left[\rho\left(\frac{\mbf{v}^2}{2} + e_{\rm g}
  + \Phi\right) + \frac{\mbf{B}^2}{8\pi}\right]
  + \mbf{\nabla}\cdot\left[
    \rho \mbf{v} \left(\frac{\mbf{v}^2}{2} + e_{\rm g} + p_{\rm g} + \Phi\right)
    - \frac{1}{4\pi}(\mbf{v}\times \mbf{B})\times \mbf{B}\right] = 0.
  \label{eq:totengMHDgen}
\end{equation}

Assuming 1D approximation along $s$ coordinate ($s=z$ in our calculations),
Equation (\ref{eq:totengMHDgen}) is rewritten as
$$
\hspace{-0.5cm}
\frac{\partial}{\partial t}\left[\rho\left(\frac{v_s^2}{2} + e_{\rm g}
    + \Phi\right) + \frac{B_s^2}{8\pi} + \frac{1}{2}\rho v_{\perp}^2
    + \frac{B_{\perp}^2}{8\pi}\right] 
+ \frac{1}{A}\frac{\partial}{\partial s}
\left[A\rho v_s\left(\frac{v_s^2}{2} + e_{\rm g} + \frac{p_{\rm g}}{\rho}
  + \Phi\right) \right.
$$
\begin{equation}  
  \left.+ A\left(\rho v_s \frac{\mbf{v}_{\perp}^2}{2}
  + \frac{1}{4\pi}v_s \mbf{B}_{\perp}^2
  - \frac{1}{4\pi} B_s \mbf{v}_{\perp}\cdot \mbf{B}_{\perp}\right)\right] = 0, 
\label{eq:totengMHDs}
\end{equation}
where the subscript $\perp$ indicates the two perpendicular components with
respect to the $s$ direction, and $A$ is the cross section of a flux tube.
If we consider \Alfvenic perturbations that propagate along the $+s$($//B_s$)
direction, the amplitudes of velocity and magnetic fields are related via
\begin{equation}
  \mbf{v}_{\perp} = -\frac{\mbf{B}_{\perp}}{4\pi}. 
\end{equation}
Based on this relation, we define energy per mass, $e_{\rm A}$, and energy flux,
$F_{\rm A}$, of \Alfvenic perturbations: 
\begin{equation}
  e_{\rm A} \equiv \frac{1}{2}\mbf{v}_{\perp}^2 + \frac{\mbf{B}_{\perp}^2}{8\pi\rho}
    = \mbf{v}_{\perp}^2 = \frac{\mbf{B}_{\perp}^2}{4\pi\rho}, 
\end{equation}
and
\begin{equation}
  F_{\rm A} = \frac{1}{2}\rho v_s \mbf{v}_{\perp}^2
  + \frac{1}{4\pi}v_s \mbf{B}_{\perp}^2
  - \frac{1}{4\pi}B_s \mbf{v}_{\perp}\cdot \mbf{B}_{\perp} = \rho e_{\rm A}\left(
  \frac{3}{2}v_s + v_{\rm A}\right),
  \label{eq:Alflux}
\end{equation}
where $v_{\rm A} = B_s/\sqrt{4\pi\rho}$.
Using $e_{\rm A}$, Equation (\ref{eq:totengMHDs}) is further rewritten as 
\begin{equation}
  \frac{\partial}{\partial t}\left[\rho\left(\frac{v_s^2}{2} + e_{\rm g}
    + \Phi\right) + \frac{B_s^2}{8\pi} + \rho e_{\rm A}\right]
  + \frac{1}{A}\frac{\partial}{\partial s}
  \left[A\rho v_s\left(\frac{v_s^2}{2} + e_{\rm g} + \frac{p_{\rm g}}{\rho}
    + \Phi\right) + A\rho e_{\rm A} \left(v_{\rm A} + \frac{3}{2}v_s \right)
    \right] = 0. 
\end{equation}
Then, we can divide Equation (\ref{eq:totengMHDs}) into the \Alfvenic part
\citep{sn05,suz08},
\begin{equation}
  \frac{\partial}{\partial t}(\rho e_{\rm A}) + \frac{1}{A}
  \frac{\partial}{\partial s}\left[A\rho e_{\rm A}\left(v_{\rm A}+\frac{3}{2}v_s
    \right)\right] - v_s\frac{\partial p_{\rm A}}{\partial s}
  = -\gamma_{\rm A}\rho e_{\rm A}
  \label{eq:Alfeng}
\end{equation}
and the hydrodynamic part,
\begin{equation}
  \frac{\partial}{\partial t}\left[\rho\left(\frac{v_s^2}{2} + e_{\rm g}
  + \Phi\right) \right] + \frac{1}{A}\frac{\partial}{\partial s}
  \left[A\rho v_s\left(\frac{v_s^2}{2} + e_{\rm g} + \frac{p_{\rm g}}{\rho}
  + \Phi\right) \right] 
  + v_s\frac{\partial p_{\rm A}}{\partial s}
=\gamma_{\rm A}\rho e_{\rm A},
\label{eq:hdreng}
\end{equation}
where $p_{\rm A}=\frac{\mbf{B}_{\perp}^2}{8\pi}=\frac{1}{2}\rho e_{\rm A}$ is
the magnetic pressure associated with \Alfvenic perturbations. 
When deriving Equation (\ref{eq:hdreng}), we use that $B_s$ is constant with
time in the 1D approximation. 
Equations (\ref{eq:Alfeng}) and (\ref{eq:hdreng}) show that both components
are connected by the momentum exchange through magnetic pressure
($v_s\frac{\partial p_{\rm A}}{\partial s}$) and the heating by the dissipation
of \Alfvenic perturbations ($\gamma_{\rm A}\rho e_{\rm A}$).
Readers may recognize that Equation (\ref{eq:Alfeng}) is the first law of
thermodynamics for \Alfvenic perturbations; the energy density,
$\rho e_{\rm A}$, changes by the difference between the incoming and outgoing
energy fluxes (2nd term on the left-hand side.),
the work done on gas (3rd term on the left-hand side.) and dissipation
(the right-hand side). 
Transforming the Eulerian form of Equation (\ref{eq:Alfeng}) to the Lagrangian
form, we get Equation (\ref{eq:alfeng}), which we are solving in our
calculations: 
\begin{equation}
  \rho \frac{d e_{\rm A}}{dt} + \frac{\partial}{\partial s}
  \left[A\rho e_{\rm A}\left(v_{\rm A} + \frac{1}{2}v_s\right)\right]
  - v_s\frac{\partial p_{\rm A}}{\partial s} 
  = -\gamma_{\rm A}\rho e_{\rm A}
\end{equation}

The second and third terms on the right-hand side of Equation (\ref{eq:Alfeng})
can be combined to give \citep{jac77}
\begin{equation}
  \frac{\partial}{\partial t}(\rho e_{\rm A}) + \frac{v_{\rm A}}{v_{\rm A} + v_s}
  \frac{1}{A}\frac{\partial}{\partial s}\left[A\rho e_{\rm A}
    \frac{(v_{\rm A}+v_s)^2}{v_{\rm A}}\right] 
  = -\gamma_{\rm A}\rho e_{\rm A}.
  \label{eq:waveaction}
\end{equation}
This equation indicates that, for undamped ($\gamma_{\rm A}=0$) \Alfven waves
under time-steady condition ($\frac{\partial}{\partial t}=0$), 
$\rho e_{\rm A}\frac{(v_{\rm A}+v_s)^2}{v_{\rm A}}$ is an adiabatic constant,
which is called wave action. 

In this paper, we assume that waves with different wave frequencies evolve
in an independent manner, and we can derive Equation (\ref{eq:alfeng})
for wave energy density, $\tilde{e}_{\rm A}(\omega)$, at $\omega$.  
In order to derive Equation (\ref{eq:alfidx}),
we here explicitly write an equation for $\tilde{e}_{\rm A}(\omega)$ by using
Equation (\ref{eq:wvengindeg}),
\begin{equation}
  \rho\frac{d}{dt}\left(\tilde{e}_{\rm A}(\omega_0)f(\omega/\omega_0)\right)
  + \frac{1}{A}\frac{\partial}{\partial z}\left[A \rho \tilde{e}_{\rm A}
  (\omega_0)f(\omega/\omega_0)\left(v_{\rm A}+\frac{v_z}{2}\right)\right]
  -\frac{v_z}{2}\frac{\partial}{\partial z}\left[\rho \tilde{e}_{\rm A}
    (\omega_0)f(\omega/\omega_0)\right] = -\gamma_{\rm A}\rho
  \tilde{e}_{\rm A}(\omega_0)f(\omega/\omega_0), 
\end{equation}
which is further transformed as
$$
f(\omega/\omega_0)\left[\rho \frac{d\tilde{e}_{\rm A}(\omega_0)}{dt}
  + \frac{1}{A}\frac{\partial}{\partial z}\left\{A \rho
  \tilde{e}_{\rm A}(\omega_0)\left(v_{\rm A} + \frac{1}{2}v_z\right)\right\}
  - \frac{v_z}{2}\frac{\partial}{\partial z}\left\{\rho
  \tilde{e}_{\rm A}(\omega_0)\right\}\right] + \rho \tilde{e}_{\rm A}(\omega_0)
\left[\frac{d}{dt}f(\omega/\omega_0) + v_{\rm A}\frac{\partial}{\partial z}
  f(\omega/\omega_0)\right]
$$
\begin{equation}
  = -\gamma_{\rm A} \rho\tilde{e}_{\rm A}(\omega_0)f(\omega/\omega_0).
  \label{eq:alfengwdbnd}
\end{equation}
We substitute the equation at $\omega=\omega_0$,
\begin{equation}
  \rho \frac{d\tilde{e}_{\rm A}(\omega_0)}{dt} + \frac{1}{A}
\frac{\partial}{\partial z}\left[A \rho \tilde{e}_{\rm A}(\omega_0)
  \left(v_{\rm A} + \frac{1}{2}v_z\right)\right]
- \frac{v_z}{2}\frac{\partial}{\partial z}\left\{\rho
\tilde{e}_{\rm A}(\omega_0)\right\}
= - \gamma_{\rm A}\rho \tilde{e}_{\rm A}(\omega_0), 
\end{equation}
into Equation (\ref{eq:alfengwdbnd}), and we obtain
\begin{equation}
  \frac{d}{dt}f(\omega/\omega_0) + v_{\rm A}\frac{\partial}{\partial z}
  f(\omega/\omega_0) = \left(\gamma_{\rm A}(\omega_0) - \gamma_{\rm A}(\omega)
  \right) f(\omega/\omega_0). 
\end{equation}
If $f(\omega/\omega_0)$ has a power-law dependence, Equation (\ref{eq:fomega}), we can derive Equation (\ref{eq:alfidx}).

\end{appendix}


\begin{thebibliography}{84}
\expandafter\ifx\csname natexlab\endcsname\relax\def\natexlab#1{#1}\fi

\bibitem[{{Ackermann} {et~al.}(2014){Ackermann}, {Albert}, {Atwood}, {Baldini},
  {Ballet}, {Barbiellini}, {Bastieri}, {Bellazzini}, {Bissaldi}, {Blandford},
  {Bloom}, {Bottacini}, {Brandt}, {Bregeon}, {Bruel}, {Buehler}, {Buson},
  {Caliandro}, {Cameron}, {Caragiulo}, {Caraveo}, {Cavazzuti}, {Cecchi},
  {Charles}, {Chekhtman}, {Chiang}, {Chiaro}, {Ciprini}, {Claus},
  {Cohen-Tanugi}, {Conrad}, {Cutini}, {D'Ammando}, {de Angelis}, {de Palma},
  {Dermer}, {Digel}, {Di Venere}, {Silva}, {Drell}, {Favuzzi}, {Ferrara},
  {Focke}, {Franckowiak}, {Fukazawa}, {Funk}, {Fusco}, {Gargano}, {Gasparrini},
  {Germani}, {Giglietto}, {Giordano}, {Giroletti}, {Godfrey}, {Gomez-Vargas},
  {Grenier}, {Guiriec}, {Hadasch}, {Harding}, {Hays}, {Hewitt}, {Hou},
  {Jogler}, {J{\'o}hannesson}, {Johnson}, {Johnson}, {Kamae}, {Kataoka},
  {Kn{\"o}dlseder}, {Kocevski}, {Kuss}, {Larsson}, {Latronico}, {Longo},
  {Loparco}, {Lovellette}, {Lubrano}, {Malyshev}, {Manfreda}, {Massaro},
  {Mayer}, {Mazziotta}, {McEnery}, {Michelson}, {Mitthumsiri}, {Mizuno},
  {Monzani}, {Morselli}, {Moskalenko}, {Murgia}, {Nemmen}, {Nuss}, {Ohsugi},
  {Omodei}, {Orienti}, {Orlando}, {Ormes}, {Paneque}, {Panetta}, {Perkins},
  {Pesce-Rollins}, {Petrosian}, {Piron}, {Pivato}, {Rain{\`o}}, {Rando},
  {Razzano}, {Razzaque}, {Reimer}, {Reimer}, {S{\'a}nchez-Conde}, {Schaal},
  {Schulz}, {Sgr{\`o}}, {Siskind}, {Spandre}, {Spinelli}, {Stawarz}, {Strong},
  {Suson}, {Tahara}, {Takahashi}, {Thayer}, {Tibaldo}, {Tinivella}, {Torres},
  {Tosti}, {Troja}, {Uchiyama}, {Vianello}, {Werner}, {Winer}, {Wood}, {Wood},
  \& {Zaharijas}}]{ack14}
{Ackermann}, M., {Albert}, A., {Atwood}, W.~B., {et~al.} 2014, \apj, 793, 64

\bibitem[{{Alazraki} \& {Couturier}(1971)}]{ac71}
{Alazraki}, G., \& {Couturier}, P. 1971, \aap, 13, 380

\bibitem[{{Balbus} \& {Hawley}(1991)}]{bh91}
{Balbus}, S.~A., \& {Hawley}, J.~F. 1991, \apj, 376, 214

\bibitem[{{Barkov} \& {Bosch-Ramon}(2014)}]{bb14}
{Barkov}, M.~V., \& {Bosch-Ramon}, V. 2014, \aap, 565, A65

\bibitem[{{Belcher}(1971)}]{bel71}
{Belcher}, J.~W. 1971, \apj, 168, 509

\bibitem[{{Beresnyak} \& {Lazarian}(2008)}]{bl08}
{Beresnyak}, A., \& {Lazarian}, A. 2008, \apj, 678, 961

\bibitem[{{Berkhuijsen} {et~al.}(1971){Berkhuijsen}, {Haslam}, \&
  {Salter}}]{ber71}
{Berkhuijsen}, E.~M., {Haslam}, C.~G.~T., \& {Salter}, C.~J. 1971, \aap, 14,
  252

\bibitem[{{Bland-Hawthorn} \& {Cohen}(2003)}]{bh03}
{Bland-Hawthorn}, J., \& {Cohen}, M. 2003, \apj, 582, 246

\bibitem[{{Bland-Hawthorn} {et~al.}(2013){Bland-Hawthorn}, {Maloney},
  {Sutherland}, \& {Madsen}}]{bla13}
{Bland-Hawthorn}, J., {Maloney}, P.~R., {Sutherland}, R.~S., \& {Madsen}, G.~J.
  2013, \apj, 778, 58

\bibitem[{{Breitschwerdt} {et~al.}(1991){Breitschwerdt}, {McKenzie}, \&
  {Voelk}}]{bre91}
{Breitschwerdt}, D., {McKenzie}, J.~F., \& {Voelk}, H.~J. 1991, \aap, 245, 79

\bibitem[{{Carretti} {et~al.}(2013){Carretti}, {Crocker}, {Staveley-Smith},
  {Haverkorn}, {Purcell}, {Gaensler}, {Bernardi}, {Kesteven}, \&
  {Poppi}}]{car13}
{Carretti}, E., {Crocker}, R.~M., {Staveley-Smith}, L., {et~al.} 2013, \nat,
  493, 66

\bibitem[{{Chandrasekhar}(1961)}]{cha61}
{Chandrasekhar}, S. 1961, {Hydrodynamic and hydromagnetic stability} (Oxford:
  Clarendon)

\bibitem[{{Chandrasekhar} \& {Fermi}(1953)}]{cf53}
{Chandrasekhar}, S., \& {Fermi}, E. 1953, \apj, 118, 113

\bibitem[{{Cheng} {et~al.}(2015){Cheng}, {Chernyshov}, {Dogiel}, \&
  {Ko}}]{che15}
{Cheng}, K.~S., {Chernyshov}, D.~O., {Dogiel}, V.~A., \& {Ko}, C.~M. 2015,
  \apj, 804, 135

\bibitem[{{Cho} \& {Lazarian}(2003)}]{cl03}
{Cho}, J., \& {Lazarian}, A. 2003, \mnras, 345, 325

\bibitem[{{Crocker}(2012)}]{cro12}
{Crocker}, R.~M. 2012, \mnras, 423, 3512

\bibitem[{{Crocker} \& {Aharonian}(2011)}]{ca11}
{Crocker}, R.~M., \& {Aharonian}, F. 2011, Physical Review Letters, 106, 101102

\bibitem[{{Crocker} {et~al.}(2015){Crocker}, {Bicknell}, {Taylor}, \&
  {Carretti}}]{cro15}
{Crocker}, R.~M., {Bicknell}, G.~V., {Taylor}, A.~M., \& {Carretti}, E. 2015,
  \apj, 808, 107

\bibitem[{{Crocker} {et~al.}(2010){Crocker}, {Jones}, {Melia}, {Ott}, \&
  {Protheroe}}]{cro10}
{Crocker}, R.~M., {Jones}, D.~I., {Melia}, F., {Ott}, J., \& {Protheroe}, R.~J.
  2010, \nat, 463, 65

\bibitem[{{Dobler} \& {Finkbeiner}(2008)}]{df08}
{Dobler}, G., \& {Finkbeiner}, D.~P. 2008, \apj, 680, 1222

\bibitem[{{Dobler} {et~al.}(2010){Dobler}, {Finkbeiner}, {Cholis}, {Slatyer},
  \& {Weiner}}]{dob10}
{Dobler}, G., {Finkbeiner}, D.~P., {Cholis}, I., {Slatyer}, T., \& {Weiner}, N.
  2010, \apj, 717, 825

\bibitem[{{Egger} \& {Aschenbach}(1995)}]{ea95}
{Egger}, R.~J., \& {Aschenbach}, B. 1995, \aap, 294, L25

\bibitem[{{Enokiya} {et~al.}(2014){Enokiya}, {Torii}, {Schultheis}, {Asahina},
  {Matsumoto}, {Furuhashi}, {Nakamura}, {Dobashi}, {Yoshiike}, {Sato},
  {Furukawa}, {Moribe}, {Ohama}, {Sano}, {Okamoto}, {Mori}, {Hanaoka},
  {Nishimura}, {Hayakawa}, {Okuda}, {Yamamoto}, {Kawamura}, {Mizuno}, {Onishi},
  {Morris}, \& {Fukui}}]{eno14}
{Enokiya}, R., {Torii}, K., {Schultheis}, M., {et~al.} 2014, \apj, 780, 72

\bibitem[{{Everett} {et~al.}(2008){Everett}, {Zweibel}, {Benjamin}, {McCammon},
  {Rocks}, \& {Gallagher}}]{eve08}
{Everett}, J.~E., {Zweibel}, E.~G., {Benjamin}, R.~A., {et~al.} 2008, \apj,
  674, 258

\bibitem[{{Farmer} \& {Goldreich}(2004)}]{fg04}
{Farmer}, A.~J., \& {Goldreich}, P. 2004, \apj, 604, 671

\bibitem[{{Finkbeiner}(2004)}]{fin04}
{Finkbeiner}, D.~P. 2004, \apj, 614, 186

\bibitem[{{Fujita} {et~al.}(2013){Fujita}, {Ohira}, \& {Yamazaki}}]{fuj13}
{Fujita}, Y., {Ohira}, Y., \& {Yamazaki}, R. 2013, \apjl, 775, L20

\bibitem[{{Fukui} {et~al.}(2006){Fukui}, {Yamamoto}, {Fujishita}, {Kudo},
  {Torii}, {Nozawa}, {Takahashi}, {Matsumoto}, {Machida}, {Kawamura},
  {Yonekura}, {Mizuno}, {Onishi}, \& {Mizuno}}]{fuk06}
{Fukui}, Y., {Yamamoto}, H., {Fujishita}, M., {et~al.} 2006, Science, 314, 106

\bibitem[{{Genzel} {et~al.}(2010){Genzel}, {Eisenhauer}, \&
  {Gillessen}}]{gen10}
{Genzel}, R., {Eisenhauer}, F., \& {Gillessen}, S. 2010, Reviews of Modern
  Physics, 82, 3121

\bibitem[{{Goldreich} \& {Sridhar}(1995)}]{gs95}
{Goldreich}, P., \& {Sridhar}, S. 1995, \apj, 438, 763

\bibitem[{{Guo} \& {Mathews}(2012)}]{gm12}
{Guo}, F., \& {Mathews}, W.~G. 2012, \apj, 756, 181

\bibitem[{{Inoue}(2012)}]{ino12}
{Inoue}, T. 2012, \apj, 760, 43

\bibitem[{{Jacques}(1977)}]{jac77}
{Jacques}, S.~A. 1977, \apj, 215, 942

\bibitem[{{Kataoka} {et~al.}(2015){Kataoka}, {Tahara}, {Totani}, {Sofue},
  {Inoue}, {Nakashima}, \& {Cheung}}]{kat15}
{Kataoka}, J., {Tahara}, M., {Totani}, T., {et~al.} 2015, \apj, 807, 77

\bibitem[{{Kataoka} {et~al.}(2013){Kataoka}, {Tahara}, {Totani}, {Sofue},
  {Stawarz}, {Takahashi}, {Takeuchi}, {Tsunemi}, {Kimura}, {Takei}, {Cheung},
  {Inoue}, \& {Nakamori}}]{kat13}
---. 2013, \apj, 779, 57

\bibitem[{{Kopp} \& {Holzer}(1976)}]{kh76}
{Kopp}, R.~A., \& {Holzer}, T.~E. 1976, \solphys, 49, 43

\bibitem[{{Koyama} {et~al.}(1996){Koyama}, {Maeda}, {Sonobe}, {Takeshima},
  {Tanaka}, \& {Yamauchi}}]{koy96}
{Koyama}, K., {Maeda}, Y., {Sonobe}, T., {et~al.} 1996, \pasj, 48, 249

\bibitem[{{Kulsrud} \& {Pearce}(1969)}]{kp69}
{Kulsrud}, R., \& {Pearce}, W.~P. 1969, \apj, 156, 445

\bibitem[{{Kuzio de Naray} {et~al.}(2009){Kuzio de Naray}, {McGaugh}, \&
  {Mihos}}]{kuz09}
{Kuzio de Naray}, R., {McGaugh}, S.~S., \& {Mihos}, J.~C. 2009, \apj, 692, 1321

\bibitem[{{Lacki}(2014)}]{lac14}
{Lacki}, B.~C. 2014, \mnras, 444, L39

\bibitem[{{Lamers} \& {Cassinelli}(1999)}]{lc99}
{Lamers}, H.~J.~G.~L.~M., \& {Cassinelli}, J.~P. 1999, {Introduction to Stellar
  Winds}, 452

\bibitem[{{Lazarian}(2016)}]{laz16}
{Lazarian}, A. 2016, \apj, 833, 131

\bibitem[{{Machida} {et~al.}(2013){Machida}, {Nakamura}, {Kudoh}, {Akahori},
  {Sofue}, \& {Matsumoto}}]{mac13}
{Machida}, M., {Nakamura}, K.~E., {Kudoh}, T., {et~al.} 2013, \apj, 764, 81

\bibitem[{{Machida} {et~al.}(2009){Machida}, {Matsumoto}, {Nozawa},
  {Takahashi}, {Fukui}, {Kudo}, {Torii}, {Yamamoto}, {Fujishita}, \&
  {Tomisaki}}]{mac09}
{Machida}, M., {Matsumoto}, R., {Nozawa}, S., {et~al.} 2009, \pasj, 61, 411

\bibitem[{{Matsumoto} \& {Suzuki}(2012)}]{ms12}
{Matsumoto}, T., \& {Suzuki}, T.~K. 2012, \apj, 749, 8

\bibitem[{{Matsumoto} \& {Suzuki}(2014)}]{ms14}
---. 2014, \mnras, 440, 971

\bibitem[{{Mertsch} \& {Sarkar}(2011)}]{ms11}
{Mertsch}, P., \& {Sarkar}, S. 2011, Physical Review Letters, 107, 091101

\bibitem[{{Miller} \& {Bregman}(2016)}]{mb16}
{Miller}, M.~J., \& {Bregman}, J.~N. 2016, \apj, 829, 9

\bibitem[{{Miyamoto} \& {Nagai}(1975)}]{mn75}
{Miyamoto}, M., \& {Nagai}, R. 1975, \pasj, 27, 533

\bibitem[{{Miyamoto} {et~al.}(2014){Miyamoto}, {Imamura}, {Tokumaru}, {Ando},
  {Isobe}, {Asai}, {Shiota}, {Toda}, {H{\"a}usler}, {P{\"a}tzold}, {Nabatov},
  \& {Nakamura}}]{miy14}
{Miyamoto}, M., {Imamura}, T., {Tokumaru}, M., {et~al.} 2014, \apj, 797, 51

\bibitem[{{Mou} {et~al.}(2015){Mou}, {Yuan}, {Gan}, \& {Sun}}]{mou15}
{Mou}, G., {Yuan}, F., {Gan}, Z., \& {Sun}, M. 2015, \apj, 811, 37

\bibitem[{{Navarro} {et~al.}(1996){Navarro}, {Frenk}, \& {White}}]{nav96}
{Navarro}, J.~F., {Frenk}, C.~S., \& {White}, S.~D.~M. 1996, \apj, 462, 563

\bibitem[{{Parker}(1966)}]{pak66}
{Parker}, E.~N. 1966, \apj, 145, 811

\bibitem[{{Pillai} {et~al.}(2015){Pillai}, {Kauffmann}, {Tan}, {Goldsmith},
  {Carey}, \& {Menten}}]{pil15}
{Pillai}, T., {Kauffmann}, J., {Tan}, J.~C., {et~al.} 2015, \apj, 799, 74

\bibitem[{{Planck Collaboration} {et~al.}(2013){Planck Collaboration}, {Ade},
  {Aghanim}, {Arnaud}, {Ashdown}, {Atrio-Barandela}, {Aumont}, {Baccigalupi},
  {Balbi}, {Banday}, {Barreiro}, {Bartlett}, {Battaner}, {Benabed},
  {Beno{\^i}t}, {Bernard}, {Bersanelli}, {Bonaldi}, {Bond}, {Borrill},
  {Bouchet}, {Burigana}, {Cabella}, {Cardoso}, {Catalano}, {Cay{\'o}n},
  {Chary}, {Chiang}, {Christensen}, {Clements}, {Colombo}, {Coulais}, {Crill},
  {Cuttaia}, {Danese}, {D'Arcangelo}, {Davis}, {de Bernardis}, {de Gasperis},
  {de Rosa}, {de Zotti}, {Delabrouille}, {Dickinson}, {Diego}, {Dobler},
  {Dole}, {Donzelli}, {Dor{\'e}}, {D{\"o}rl}, {Douspis}, {Dupac}, {Efstathiou},
  {En{\ss}lin}, {Eriksen}, {Finelli}, {Forni}, {Frailis}, {Franceschi},
  {Galeotta}, {Ganga}, {Giard}, {Giardino}, {Gonz{\'a}lez-Nuevo}, {G{\'o}rski},
  {Gratton}, {Gregorio}, {Gruppuso}, {Hansen}, {Harrison}, {Helou},
  {Henrot-Versill{\'e}}, {Hern{\'a}ndez-Monteagudo}, {Hildebrandt}, {Hivon},
  {Hobson}, {Holmes}, {Hornstrup}, {Hovest}, {Huffenberger}, {Jaffe},
  {Jagemann}, {Jewell}, {Jones}, {Juvela}, {Keih{\"a}nen}, {Knoche}, {Knox},
  {Kunz}, {Kurki-Suonio}, {Lagache}, {L{\"a}hteenm{\"a}ki}, {Lamarre},
  {Lasenby}, {Lawrence}, {Leach}, {Leonardi}, {Lilje}, {Linden-V{\o}rnle},
  {L{\'o}pez-Caniego}, {Lubin}, {Mac{\'{\i}}as-P{\'e}rez}, {Maffei}, {Maino},
  {Mandolesi}, {Maris}, {Marshall}, {Martin}, {Mart{\'{\i}}nez-Gonz{\'a}lez},
  {Masi}, {Massardi}, {Matarrese}, {Matthai}, {Mazzotta}, {Meinhold},
  {Melchiorri}, {Mendes}, {Mennella}, {Mitra}, {Moneti}, {Montier}, {Morgante},
  {Munshi}, {Murphy}, {Naselsky}, {Natoli}, {N{\o}rgaard-Nielsen}, {Noviello},
  {Novikov}, {Novikov}, {Osborne}, {Pajot}, {Paladini}, {Paoletti},
  {Partridge}, {Pearson}, {Perdereau}, {Perrotta}, {Piacentini}, {Piat},
  {Pierpaoli}, {Pietrobon}, {Plaszczynski}, {Pointecouteau}, {Polenta},
  {Ponthieu}, {Popa}, {Poutanen}, {Pratt}, {Prunet}, {Puget}, {Rachen},
  {Rebolo}, {Reinecke}, {Renault}, {Ricciardi}, {Riller}, {Ristorcelli},
  {Rocha}, {Rosset}, {Rubi{\~n}o-Mart{\'{\i}}n}, {Rusholme}, {Sandri},
  {Savini}, {Schaefer}, {Scott}, {Smoot}, {Spencer}, {Stivoli}, {Sudiwala},
  {Suur-Uski}, {Sygnet}, {Tauber}, {Terenzi}, {Toffolatti}, {Tomasi},
  {Tristram}, {T{\"u}rler}, {Umana}, {Valenziano}, {Van Tent}, {Vielva},
  {Villa}, {Vittorio}, {Wade}, {Wandelt}, {White}, {Yvon}, {Zacchei}, \&
  {Zonca}}]{pla13}
{Planck Collaboration}, {Ade}, P.~A.~R., {Aghanim}, N., {et~al.} 2013, \aap,
  554, A139

\bibitem[{{Ptuskin} {et~al.}(1997){Ptuskin}, {Voelk}, {Zirakashvili}, \&
  {Breitschwerdt}}]{ptu97}
{Ptuskin}, V.~S., {Voelk}, H.~J., {Zirakashvili}, V.~N., \& {Breitschwerdt}, D.
  1997, \aap, 321, 434

\bibitem[{{Sano} {et~al.}(1999){Sano}, {Inutsuka}, \& {Miyama}}]{san99}
{Sano}, T., {Inutsuka}, S., \& {Miyama}, S.~M. 1999, in Astrophysics and Space
  Science Library, Vol. 240, Numerical Astrophysics, ed. S.~M. {Miyama},
  K.~{Tomisaka}, \& T.~{Hanawa}, 383

\bibitem[{{Sasaki} {et~al.}(2015){Sasaki}, {Asano}, \& {Terasawa}}]{sas15}
{Sasaki}, K., {Asano}, K., \& {Terasawa}, T. 2015, \apj, 814, 93

\bibitem[{{Shoda} \& {Yokoyama}(2016)}]{sy16}
{Shoda}, M., \& {Yokoyama}, T. 2016, \apj, 820, 123

\bibitem[{{Snowden} {et~al.}(1997){Snowden}, {Egger}, {Freyberg}, {McCammon},
  {Plucinsky}, {Sanders}, {Schmitt}, {Tr{\"u}mper}, \& {Voges}}]{sno97}
{Snowden}, S.~L., {Egger}, R., {Freyberg}, M.~J., {et~al.} 1997, \apj, 485, 125

\bibitem[{{Sofue}(1977)}]{sof77}
{Sofue}, Y. 1977, \aap, 60, 327

\bibitem[{{Sofue}(2000)}]{sof00}
---. 2000, \apj, 540, 224

\bibitem[{{Sofue}(2015)}]{sof15}
---. 2015, \pasj, 67, 75

\bibitem[{{Sofue} \& {Handa}(1984)}]{sh84}
{Sofue}, Y., \& {Handa}, T. 1984, \nat, 310, 568

\bibitem[{{Su} {et~al.}(2010){Su}, {Slatyer}, \& {Finkbeiner}}]{su10}
{Su}, M., {Slatyer}, T.~R., \& {Finkbeiner}, D.~P. 2010, \apj, 724, 1044

\bibitem[{{Suzuki}(2004)}]{suz04}
{Suzuki}, T.~K. 2004, \mnras, 349, 1227

\bibitem[{{Suzuki} {et~al.}(2015){Suzuki}, {Fukui}, {Torii}, {Machida}, \&
  {Matsumoto}}]{suz15}
{Suzuki}, T.~K., {Fukui}, Y., {Torii}, K., {Machida}, M., \& {Matsumoto}, R.
  2015, \mnras, 454, 3049

\bibitem[{{Suzuki} \& {Inutsuka}(2005)}]{si05}
{Suzuki}, T.~K., \& {Inutsuka}, S.-i. 2005, \apjl, 632, L49

\bibitem[{{Suzuki} \& {Inutsuka}(2006)}]{si06}
---. 2006, Journal of Geophysical Research (Space Physics), 111, 6101

\bibitem[{{Suzuki} \& {Inutsuka}(2009)}]{si09}
---. 2009, \apjl, 691, L49

\bibitem[{{Suzuki} \& {Inutsuka}(2014)}]{si14}
---. 2014, \apj, 784, 121

\bibitem[{{Suzuki} \& {Nagataki}(2005)}]{sn05}
{Suzuki}, T.~K., \& {Nagataki}, S. 2005, \apj, 628, 914

\bibitem[{{Suzuki} {et~al.}(2008){Suzuki}, {Sumiyoshi}, \& {Yamada}}]{suz08}
{Suzuki}, T.~K., {Sumiyoshi}, K., \& {Yamada}, S. 2008, \apj, 678, 1200

\bibitem[{{Torii} {et~al.}(2010{\natexlab{a}}){Torii}, {Kudo}, {Fujishita},
  {Kawase}, {Yamamoto}, {Kawamura}, {Mizuno}, {Onishi}, {Mizuno}, {Machida},
  {Takahashi}, {Nozawa}, {Matsumoto}, \& {Fukui}}]{tor10b}
{Torii}, K., {Kudo}, N., {Fujishita}, M., {et~al.} 2010{\natexlab{a}}, \pasj,
  62, 1307

\bibitem[{{Torii} {et~al.}(2010{\natexlab{b}}){Torii}, {Kudo}, {Fujishita},
  {Kawase}, {Okuda}, {Yamamoto}, {Kawamura}, {Mizuno}, {Onishi}, {Machida},
  {Takahashi}, {Nozawa}, {Matsumoto}, {Ott}, {Tanaka}, {Yamaguchi}, {Ezawa},
  {Stutzki}, {Bertoldi}, {Koo}, {Bronfman}, {Burton}, {Benz}, {Ogawa}, \&
  {Fukui}}]{tor10a}
---. 2010{\natexlab{b}}, \pasj, 62, 675

\bibitem[{{Veilleux} {et~al.}(2005){Veilleux}, {Cecil}, \&
  {Bland-Hawthorn}}]{vei05}
{Veilleux}, S., {Cecil}, G., \& {Bland-Hawthorn}, J. 2005, \araa, 43, 769

\bibitem[{{Velikhov}(1959)}]{vel59}
{Velikhov}, E.~P. 1959, Zh. Eksp. Teor. Fiz., 36, 1398

\bibitem[{{Verdini} {et~al.}(2012){Verdini}, {Grappin}, {Pinto}, \&
  {Velli}}]{ver12}
{Verdini}, A., {Grappin}, R., {Pinto}, R., \& {Velli}, M. 2012, \apjl, 750, L33

\bibitem[{{Vishniac} \& {Cho}(2001)}]{vc01}
{Vishniac}, E.~T., \& {Cho}, J. 2001, \apj, 550, 752

\bibitem[{{Vishniac} \& {Shapovalov}(2014)}]{vs14}
{Vishniac}, E.~T., \& {Shapovalov}, D. 2014, \apj, 780, 144

\bibitem[{{Wentzel}(1968)}]{wen68}
{Wentzel}, D.~G. 1968, \apj, 152, 987

\bibitem[{{Yan} \& {Lazarian}(2002)}]{yl02}
{Yan}, H., \& {Lazarian}, A. 2002, Physical Review Letters, 89, 281102

\bibitem[{{Zirakashvili} {et~al.}(1996){Zirakashvili}, {Breitschwerdt},
  {Ptuskin}, \& {Voelk}}]{zir96}
{Zirakashvili}, V.~N., {Breitschwerdt}, D., {Ptuskin}, V.~S., \& {Voelk}, H.~J.
  1996, \aap, 311, 113

\bibitem[{{Zubovas} {et~al.}(2011){Zubovas}, {King}, \& {Nayakshin}}]{zub11}
{Zubovas}, K., {King}, A.~R., \& {Nayakshin}, S. 2011, \mnras, 415, L21

\end{thebibliography}


\end{document}